\documentclass[12pt]{article}
\usepackage[dvips]{graphicx}

\usepackage{color}
\usepackage[x11names,rgb]{xcolor}

\usepackage{authblk}

\usepackage{tikz}
\usetikzlibrary{arrows,snakes,backgrounds}

\usepackage{amsfonts}
\usepackage{amsmath}
\usepackage{amssymb}
\usepackage{amsthm}
\usepackage{enumerate}
\usepackage{fancybox}

\usepackage{natbib}
\bibpunct{(}{)}{;}{a}{,}{,}
\usepackage[pdftex]{hyperref}
\hypersetup{colorlinks,%
            citecolor=blue,%
            filecolor=blue,
            linkcolor=blue,%
            urlcolor=blue,
            }

\newtheorem{proposition}{{\bf \sc Proposition}}

\newtheorem{lemma}[proposition]{{\bf \sc Lemma}}

\newtheorem{corollary}[proposition]{{\bf \sc Corollary}}
\newtheorem{definition}{{\bf \sc Definition}}
\newtheorem{remark}{{\bf \sc Remark}}

\newtheorem{claim}{{\bf \sc Claim}}

\theoremstyle{remark} 
\def\eproof{\qed}

{\begin{Sbox}\begin{minipage}}%
{\end{minipage}\end{Sbox}\fbox{\TheSbox}}

\newenvironment{revs}
  {\begin{color}{black} \ignorespaces} 
 {\end{color}}

\begin{document}

\title{Bring a friend! Privately or Publicly?\thanks{%
We wish to thank Department Editor Juanjuan Zhang, an anonymous associate editor and two anonymous referees of Management Science for their precious suggestions.   We  acknowledge Gani Aldashev, Jean Marie Baland, Paul Belleflamme, Ennio Bilancini, Kostas Bimpikis, Marc Bourreau, Timoteo Carletti, Jacques Cr\'{e}mer, Rosa Branca Esteves, Stefano Galavotti, Andrea Galeotti,  Edoardo Grillo, Mathias Hungerbuhler,   Mark Le Quement, Andrea Mantovani, Larry Samuelson, K\'{a}roly Tak\'{a}cs, Eric Toulemonde,
 the audience to the seminars in  Central European University, Cergy-Pontoise, Corvinus University, FUSL, Modena, MTA TK ``L\"{e}ndulet'' RECENS, Sassari, University  of East Anglia, Bologna, IMT-Lucca,
and the the participants at the AFSE 2014, ASSET 2013 and ASSET 2014, Ecore Summer School 2013 (Leuven), Bomopav 2015 (Modena),  IO in the Digital Economy Workshops (Li\`{e}ge) and  GRASS 2016  for their useful comments and critics.   We also thank Lucy Scioscia for editorial assistance. Elias Carroni acknowledges the  ``Programma Master \& Back - Regione Autonoma della Sardegna" and the Labex MMD-II for financial support. Simone Righi acknowledges the ``International Mobility Fund",  ``International Publication Fund",  the ``Lend\"{u}let'' Program of the Hungarian Academy of Sciences,  the Hungarian Scientific Research Fund (OTKA K 112929) and the European Research Council (ERC) under the European Union’s Horizon 2020 research and innovation programme (grant agreement No 648693) for financial support. 
P.\ Pin acknowledges funding from the Italian Ministry of Education Progetti di Rilevante Interesse Nazionale (PRIN) grant 2015592CTH.
A previous working paper version of this paper is in \cite{carpinrighi2017}.}
}

\author[a]{Elias Carroni}
\author[b]{Paolo Pin}
\author[c,d]{Simone Righi}
\affil[a]{Department of  Economics (DSE), Alma Mater Studiorum -- Universit\`a di Bologna, Italy.
Email: \url{elias.carroni@unibo.it}}
\affil[b]{Department of Decision Sciences, IGIER and BIDSA, Bocconi, Italy.  Email: \url{paolo.pin@unibocconi.it}}
\affil[c]{Department of Computer Science, University College London, United Kingdom. Email: \url{s.righi@ucl.ac.uk} }
\affil[d]{MTA TK ``Lend\"{u}let" Research Center for Educational and Network Studies (RECENS), Hungarian Academy of Sciences.}

\date{December 2018}

\maketitle

%

\newpage

\begin{abstract}
 We study the optimal referral strategy of a seller and its relationship with the type of communication  channels among consumers.  The seller faces a partially uninformed population of consumers, interconnected through a directed social network. In the network, the seller offers rewards to informed consumers (influencers) conditional on inducing purchases by uninformed consumers (influenced).  Rewards are needed to bear a communication cost   and to induce word-of-mouth (WOM) either privately (cost-per-contact) or publicly (fixed cost to inform all friends). 
From the seller's viewpoint, eliciting  Private  WOM is more costly than eliciting Public WOM.  We investigate \emph{(i)} the incentives for the
seller to move to a denser network, inducing either Private or Public WOM and  \emph{(ii)} the optimal mix between the two types of communication.  A denser network is found to be always better, not only for information diffusion but also for seller's profits, as long as Private WOM is concerned.  Differently,  under Public WOM,  the seller  may prefer an environment with less competition between informed consumers and the presence of highly connected influencers (hubs) is the main driver to make network density beneficial to profits. When the seller is able to discriminate between Private and Public WOM, the optimal strategy is to cheaply  incentivize the more connected people to pass on the information publicly and then offer a high bonus for Private WOM.  
\end{abstract}

\noindent \textbf{JEL\ classification}: D42,  D83, D85.


\section{Introduction}
\label{sec:intro}
Programs that attribute referral bonuses to customers are an established marketing strategy through which companies are able to increase the diffusion of their products. This strategy is effective since consumers are part of a network of acquaintances, and thus can be incentivized to use their social relationships to diffuse the knowledge about the existence of a company's product. Mass media advertisements are an imperfect alternative to increase diffusion, as information derived from mass media is not fully trusted by consumers, who tend to be more influenced by their social neighbors (\citealt{lazarsfeld1955personal}).  

In a typical referral--bonus program, a company offers rewards to its established customer base, provided that they are able to convince some of their peers to become new clients. In order to obtain rewards, existing customers need to invest in their social network by informing their peers about a product. Depending on their willingness to pay, newly--informed agents will then decide whether or not to purchase.   The optimal reward structure to diffuse referrals has been studied by \cite{Biyalogorsky} and more recently by \cite{leduc2015pricing} and \cite{lobel2016customer}. These previous studies, however, overlook  the channel used for product diffusion. The use of different channels intrinsically generates different patterns of word of mouth (WOM). Modern communication technologies offer firms many alternatives such as instant messaging, chats, emails and online social networks (OSNs). In all these cases, informing the ego--network turns out to be very cheap for a consumer, but the type of communication concerned is essentially different. For example, when friends are contacted via instant messaging or emails, referrers have to inform {\it each friend} individually, in the hope of obtaining more discounts. We label this type of communication {\it Private WOM}. Differently, in online social networks (OSNs), informing one's entire ego--network is very cheap and independent from the number of people in one's network. Indeed, in this case communication among consumers entails a fixed cost so that costs per contact are negligible: just one or a few clicks is enough to inform all acquaintances. We label this type of communication {\it Public WOM}.
Moreover, the use of different channels implies different network structures and densities among which the firm has at least some decision power.  For example, people may have different friends in different social networks and   thus  the total ego-network on Facebook plus Twitter is a superset of the Facebook-only or Twitter-only ego-network   (i.e. adding one more channel can only extend the number of individual links, albeit weakly).   By adding the buttons  ``share on Facebook'' or ``share by e-mail", the company is effectively deciding the channels through which consumers will be able to pass on information and the type of WOM elicited.\footnote{%
We use the terms \emph{private} and \emph{public} to distinguish between different social networks and different costs of communication, and not directly between different types of observability.}

There are many successful examples of referral strategies in different markets. One of the most successful programs can be found in the market of online storage services, such as Dropbox, which offers free storage space to clients who bring other subscribers.   Dropbox's strategy elicits both Private and Public WOM among consumers,  offering them the possibility to invite friends by sharing a link on both social media and through private messaging.   According to \cite{huston}, founder and CEO of Dropbox, the referral program extended their client basis by 60\% in 2009 and referrals were responsible for 35\% of new daily signups proving to be, over time, pivotal for the  company's success (\citealt{berman2016referral}).
Moreover, banks offer advantageous conditions to existing customers who are able to bring on new customers,   typically through Private WOM given the nature of their services.   For example, {\it UBS}, among others, embeds the rewarding mechanism into established customer loyalty schemes that award points in exchange for referrals. These points can then be used to claim prizes (mobile phones, televisions etc.) which are typically more valuable than those offered by companies such as Dropbox.\footnote{At current prices Dropbox offers a 1 TeraByte yearly retail contract for a price of for 99\$ and a unitary referral reward of 500MB. This implies an equivalent reward per contact of  50 cents, much less than the commercial value of a mobile phone.}
Other well--known examples can be found in markets for telecommunication services (\emph{Vodafone}), television broadcasting ({\it Sky}), payment systems (such as {\it PayPal}), tourist accommodations (e.g.~{\it AirBnB}) and airlines (see the program {\it You\&Friends} launched by Turkish Airlines).

  In terms of communication channels, the seller's problem is  twofold. First, within Private or Public WOM, she needs to decide the relevant network of consumers:  should it be limited or extended as much as possible? In other terms, the issue is whether  to operate in a dense or in a sparse network. This decision is not trivial since informed clients are proposed to sustain a costly investment, but its return is uncertain due to two considerations. On one side, some of the peers contacted may not be willing to buy the product, even once they are aware of its existence. On the other side, uninformed consumers may get information about the service from multiple sources, while only one person can receive the resulting bonus. As we take these issues into account in modelling the expectations of consumers, we are able to focus on the tension between profits and product diffusion that characterizes markets with networked consumers.   Second, since giving the opportunity to share links privately or publicly elicits different incentives, a seller can further discriminate rewards based on the type of channel used to pass on the information. Thus, we study the reward structure to generate the optimal levels of Private and Public WOM. 

In this paper, a population of consumers is divided between individuals informed about the existence of the product and others who are not, with a directed network linking the former to the latter. We consider a seller providing links to current customers. The latter receives per-friend bonuses conditional on each successful referral, which is activated when the contacted friend uses the link to make a purchase.   Links can be shared with friends  publicly (posted on OSN walls), privately (shared by emails) or both. Under Public WOM, if a buyer wants to inform all her out--neighbors, she only faces a lump-sum cost of communication. Differently, under Private WOM, a cost per contact is faced.   The seller decides the size of the individual reward, as well as the price of the product, discriminating between informed and uninformed consumers. Each consumer has private information about her own characteristics (degree and preference over the product), whereas the distribution of these characteristics is common knowledge.

From the point of view of the seller, an optimal reward trades--off between unitary margins and the provision of adequate communication incentives.
Concerning Public WOM, the consumer's communication decision  depends on her popularity (out--degree): only sufficiently connected individuals are incentivized to bear the  cost of sharing a link on Facebook.  Contrastingly, the decision to pass the information in Private WOM concerns the individual link and it is not related to the size of the ego-network. In both types of WOM, reward size depends on the distribution of degrees and the in-degree drives the power of the referral program in terms of information diffusion.
 
 Our setup allows for findings along different lines. First, we are able to highlight the main trade-offs faced by a seller, given a network   distribution and a WOM setup.   Second, we study a seller's incentives to choose a more or less dense network on which to run her referral program.   Third, we analyze the optimal bonus combination when both Private and Public WOM is possible. 

We start by studying the optimal bonus when the network as well as the type of communication elicited by the referral program are given. We find that the seller sets an optimal bonus in order to target a given diffusion objective. Diffusing the same level of information would be more costly for the seller under Private WOM, as highly connected influencers help the seller in Public but not in Private WOM. In particular, in Public WOM, only highly connected individuals are incentivized to pass the information. This makes congestion in Public WOM less important for the seller  than in the private regime, where decisions are made on the single link. As a result, the seller can diffuse more information and make higher profits using Public WOM. This result provides a theoretical explanation for the observation that referral strategies, already present in the off-line world, become even more profitable and thus diffused in the OSNs world. 

We can now turn our attention to the study of  referral channels. First, we analyze the choice between operating in a sparse or dense network. Second, we characterize the optimal mix between Public and Private WOM when the seller can discriminate the bonus according to the type  of channel used to share the link.   In order to answer the first question, we consider broad classes of degree distributions and use the concept of first-order stochastic dominance (FOSD) to capture the impact of network density. In order to stimulate Private WOM, it is always better for the seller to work in a denser network. This is because, for each given bonus, the expected payment of each consumer increases as well as the proportion of people who pass on the information, with clear-cut positive impacts on profit. Differently, under public communication, the choice depends on network degree distribution. When both in-- and out--degrees are homogeneous, a seller  always has incentives to use as many channels of information diffusion as possible, thus proposing the bonus to a denser network, provided that the cost of communication is sufficiently small. Oppositely, if only the in-degrees are homogenous and/or the cost of information is sufficiently high, then the seller may prefer to offer the bonus to a sparser network. We also highlight the important role of hubs, which turn out to be the main driver toward situations in which a denser network improves both product diffusion and sellers' profits.

With regard to the optimal mix between Public and Private WOM, we  highlight 
that taking into consideration that offering a bonus incentivizes both private and public communication induces scope for discriminating strategies between the two channels. 
From the point of view of the individual consumer,  private communication turns out not to depend on her degree and it is more costly than Public WOM. As a result, the reward in the private channel is higher. 
The presence of hubs has again an important role. Indeed,  as influencing hubs become relatively more important the network becomes sparser, so that any given bonus maps into less communication. Since the Private WOM (which is non-network-based) is more expensive for the seller, the public bonus has to increase in order to move people on the public channel, thus guaranteeing optimal information diffusion.

The remaining part of this paper is divided as follows. After discussing the related literature in the following section, we outline the mathematical aspects of the model in Section \ref{model} and we perform an equilibrium analysis under Private and Public WOM in Section \ref{tradeoff}. Then, we follow up with a comparative-statics analysis on the effects of network density (Section \ref{virtual}). In Section \ref{mix} we study the optimal mix between private and public incentives to communicate, before drawing conclusions (Section \ref{conclusions}).

\section{Related literature}
\label{literature}

Word-of-mouth communication (WOM) is empirically an important phenomenon.  The seminal work of \cite{lazarsfeld1955personal} formulated the general theory that when people speak with each other and are exposed to information from media, their decisions are based on what peers and opinion leaders say rather than on what media communicate.\footnote{See also \cite{van2007new} and \cite{iyengar2011opinion} for the role of opinion leaders (or influential agents) in the diffusion of product adoptions and \cite{arndt1967role} for an early empirical study on the positive short-term sales effects of product-related conversations.}  The role of peers in decision making is especially important when we consider agents as members of a social group, so that individual behavior is influenced by local network interactions (\citealt{jackson2016economic}).\footnote{There is a growing literature on learning and diffusion in social networks, summarized in the recent surveys of  \cite{diff2016handbook} and \cite{learn2016handbook}. Diffusion of behaviours in a social network is known to depend on the connectivity distribution of the latter  and on specific features of the diffusion rule (\citealt{lopez2008diffusion}). Furthermore, the role of social influence on binary decisions for generic influence rules is studies by \cite{lopez2012influence}.} Companies' strategies, can thus account for the fact that consumers interact with neighbors (\citealt{Sundararajan,dutta}) rather than with the overall population as in the traditional network externality approach (\citealt{katz1985network}).\footnote{The concept of network locality has been used by \cite{dutta} to show the emergence of local monopolies with homogeneous firms competing in prices, by \cite{candogan2012optimal} to study optimal pricing in networks with positive network externalities, by  \cite{bloch2011pricing} to study the optimal monopoly pricing in on-line social networks and by \cite{shi2003social} to study pricing in the presence of weak and strong ties in telecommunication markets. Other important models accounting for network externalities among consumers are  \cite{fainmesser2015pricing} and \cite{Zenou}.} Along the lines of this new trend, our paper studies how to incentivize communication among consumers through a referral program, which is an alternative way to artificially create externality from consumption.\footnote{As pointed out by \citet[p. 3516]{lobel2016customer} ``A referral program provides a way to artificially create network externality inducing complementarities in friends' purchase decisions.''}

Our paper is not the first  to discuss the management of referral strategies by companies. Incentivized WOM through referrals has been studied by \cite{Biyalogorsky}, \cite{leduc2015pricing}, \cite{lobel2016customer} and \cite{kamada2017contracting}, among others.\footnote{A related strand of the literature (e.g., \citealt{GOYAL2014,BimpOzd}) links competition among companies with targeted WOM, highlighting the  marketing budget to optimize product diffusion and brand awareness.   }
\cite{Biyalogorsky} studied the optimal pricing and referral payments over consumers' lifetime. In our static model, we make consumers account for congestion  effects derived by introducing  the referral program on a network.
More similar to our model, \cite{lobel2016customer} study the optimal referral schemes in a rooted network in which a focal consumer decide whether to purchase and whether to inform friends in exchange  for a referral payment. They find that linear (e.g., ``You earn \$3 or each new friend you bring'') or threshold (e.g., ``Earn \$3 for each of the first 4 friends that purchase the product'') referral payments are a good approximation of the first best, which would be a non-monotonic payment structure (e.g., ``Bring 3 friends and earn \$20 or bring 4 and earn \$15'') hardly understandable by consumers. Relying on their quasi-optimality, we assume the linearity of payment and focus our analysis on the diffusion mechanisms (Private and/or Public WOM) and on the choice of channels (i.e., of the network density).\footnote{In some sense, limiting network density can be seen as a limit on the ego network of consumers, which is an alternative operationalization of the \cite{lobel2016customer} threshold payment.} By assuming the consumer's network as an infinite spanning tree, \cite{lobel2016customer}   do not fully take into consideration the congestion effect that is instead central in our model and in social networks characterized by high degree of clustering (\citealt{watts1998collective}).  
With a different approach,  \cite{leduc2015pricing} and \cite{kamada2017contracting} study the incentivized diffusion of products of uncertain quality.  In particular, \cite{kamada2017contracting} study free contracts as an alternative to referrals which turn out to be more efficient when relatively few people are highly interested in the product. The introduction of free contracts creates a positive externality for the referrer as it increases the probability that the information recipient (freely) adopts the product. A similar mechanism is at play in our setup. Here, the probability of adoption is lowered by the congestion, which is indirectly manipulated by the seller's choice of relevant network and channel of diffusion. Unlike the latter but similarly to our work, \cite{leduc2015pricing} embeds consumers in a network, showing that  high-degree people are worth being incentivized because even though they have more incentive to free-ride, they are powerful conduits for the diffusion of information.  Instead of fixing the network and studying the optimal referral-price pairs to diffuse information, we allow the seller to strategically choose the channel through which to implement the referral strategy.

An alternative way of looking at WOM is considering consumers as engaging in this activity even in absence of incentives.\footnote{See \cite{bloch2016handbook} for an interesting  survey on targeting and pricing in social network.} Thus, the focus of this literature is how companies can profit from boosting or reducing this activity. 
Along these lines, \cite{campbell2013word} studies the optimal pricing when few consumers are initially informed and engage in WOM. \cite{galeotti2009influencing} discuss the optimal target to maximize market penetration with WOM; \cite{galeotti2010talking} investigates the relationship between interpersonal communication and consumer investments in search and \cite{campbell2017managing} studies WOM versus advertising in a context where consumers derive utility from being perceived by peers as sources of information.  \begin{revs}Similarly to the present paper, they  take into account the fact that when too many people get informed about the product, the individual incentive
to talk about it can be dampened. Differently from us, they consider WOM as given. Under this assumption,\end{revs}  the key issue for the seller is to understand in which circumstances WOM is positive or negative for profits and then act accordingly.  Our WOM results from a deliberate incentive scheme predisposed by the seller. In other words, the strategy we analyze generates communication which would not exist otherwise.

\section{Model}
\label{model}

A seller seeks to sell a product to a measurable mass of consumers. A fraction $1-\beta$ of them is informed about the existence of seller's product.
The remaining fraction $\beta$ is not informed. 
This paper studies the seller's optimal strategies with respect to the second set of consumers.
One simple way to rationalize this focus is to assume that being or not being  informed is strongly correlated with a consumer's willingness to pay. People that are informed are also  strongly interested in the seller's product, whereas non-informed people's  reservation value  for the product is uncertain.  For the sake of simplicity, we  assume that an informed consumer  has a reservation value normalized to 1, whereas the  reservation value of a non-informed one is $v\sim U[0,1]$.\footnote{\begin{revs}The uniform distribution   as well as the normalization to $1$ are chosen to simplify computation and exposition. Qualitative results would not change with any  demand function.\end{revs} See \cite{carpinrighi2017} for an extensive analysis of the more general case.} 
All consumers are  satiated at one unit. 
The seller offers the good at  a price  $p_{1}$ to informed consumers, and at  a price $p_{2}$ to uninformed consumers.
\begin{revs}Moreover, the seller offers  to the informed customers a unitary bonus $b$ conditional on successfully informing each non-informed consumer\end{revs}.\footnote{Our viewpoint here is the study of the decision on the network chosen by the seller. In principle, other referral bonus structures could be devised.  The optimal referral schemes are studied by \cite{lobel2016customer}, fixing the price of the product and a stylized network structure.}  If $\ell$ informed  consumers pass on the information to the same uninformed buyer, and that uninformed buyer buys the product,  each of them gets $\frac{b}{\ell}$ in expected terms.
\begin{revs}
The timing of the game is as follows:
\begin{enumerate}[(i)]
\item the seller chooses $p_1$, $p_2$ and $b$;
\item informed consumers, conditional on buying the product, decide if they want to pass the information about it to the uninformed consumers;
\item uninformed consumers decide if they want to buy and, conditional on that, informed consumers get rewards from the seller.
\end{enumerate}
It is important to note that, differently from previous works such as \cite{campbell2013word} and \cite{AceBimp}, that analyzed dynamical models of WOM diffusion, our approach is based on a reduced form of diffusion that happens in a single time period.
This allows us to focus on seller's decisions given the individual incentives to communicate.
In fact, point $(ii)$ above is crucial for our analysis and it would be  overshadowed  in a dynamic setting.

In the following, we start defining  the social network on which this referral campaign happens.
Then, reducing the outcome of point $(ii)$ to the level of communication in the network,
in Section \ref{subsection:congestion} we analyze congestion effects among informed consumers.
In the first part of Section \ref{tradeoff},  we define two types of communication strategies that can happen in point $(ii)$: Private and Public WOM. 
In this way, the following of the paper analyzes the outcome of the game and the comparative statics in these two cases, considering that the seller can actually choose endogenously which one to implement.
\end{revs}

\subsection{The social network}

Informed consumers are linked with uninformed ones through a bipartite directed network, which is the only sub-network relevant for the problem at hand. While obviously any individual  can in principle have other connections, only links going from the informed  to the uniformed group can influence people's behavior. 
This network  is such that informed consumers have an out-degree $k$ distributed according to an i.i.d.~degree distribution $f(k)$, while uninformed consumers  have an in-degree $k$ distributed according to a  i.i.d.~degree distribution   $g(k)$. Nodal characteristics are private information whereas the distributions of degrees, information and reservation values are common knowledge. 
For the sake of consistency of the model,   out- and in-degrees need to match, which is equivalent to say that:
\begin{equation}
\label{eq:consistency}
(1-\beta) \sum_{k=1}^{\infty} f(k) \cdot k = \beta \sum_{k=1}^{\infty} g(k) \cdot k,\ \
\end{equation}
implying that the average in-degree (weighted  by the relative proportion of informed and uninformed  consumers) equates the average out-degree.
We define as $k_{f \min}$, $k_{f \max}$, $k_{g \min}$, $k_{g \max}$  the minimal and maximal non-null elements in the support of $f(k)$ and $g(k)$.
See also \ref{Construction}  for a description of the implicit mathematical assumptions behind this characterization.

\subsection{Communication and congestion}
\label{subsection:congestion}

If communication is costly, only some informed consumers  will pass the information on when incentivized by the bonus $b$. 
For now, let us call $L$ the fraction of links from informed to uninformed engaging in word-of-mouth communication. $L$ will depend on the cost of passing on the information, which determines how each consumer reacts to the communication  incentive provided by the bonus.
$L$ will determine both the expected fraction of uninformed consumers who receive the information and the expected value of one out-link when passing on the information, i.e., how many other agents are expected to  inform the same buyer.

On the one hand, the fraction of people receiving the information will be given by: 
 \begin{equation}
\label{eq:Gamma}
\Gamma(L) = \sum_{k=1}^{\infty} g(k) \left(  1-(1-L)^k \right) = 1 - \sum_{k=1}^{\infty} g(k) (1-L)^k,
\end{equation} 
which is an increasing and concave function of $L$, with $\Gamma(L)>L$  for interior values of $L$,\footnote{Notice that, if $g(1)<1$, then $\Gamma(L)$ is increasing and  concave in $L$, because each element $(1-L)^k$, for $k\geq2$, is decreasing and concave in $L$. Note also that, whenever $g(1)<1$ and $0 \leq L \leq 1$, we have $\Gamma(L)>L$.
$\Gamma(L)$ is actually strictly concave whenever $g(1)<1$, which is the non-trivial case of at least some receivers with multiple links.} $\Gamma(0) =0$ and $\Gamma(1)=1$.
This implies that the fraction of receivers of the word-of-mouth communication is always greater than the fraction of active communication channels, and this has a positive  effect on information diffusion. 

On the other hand, as long as the receivers have multiple in-links, the offer of the bonus also induces some competition between informed consumers. Indeed, on a social network, an uninformed individual can become aware of the product by multiple sources. Therefore, the expected value of an informed consumer when passing on  the information needs to take into consideration this congestion effect. 
Let us call  $\phi (L,p_2)$ the expected value (rescaled by $b$) of one out-link when passing the information.
Function $\phi (L,p_2)$ is given by (i)  the probability that the recipient of the message will buy the product, multiplied by  (ii) the expected bonus the sender will receive, which depends on how many other informed consumers connected to the receiver may also have passed - in expectation - the information:

\begin{eqnarray}
\phi (L,p_2) &  = & \Pr(v\geq p_2)  \sum_{k=1}^{\infty} g(k)  \left( 1 - \sum_{t=0}^{k-1} {k-1 \choose t} L^t (1-L)^{k-1-t} \frac{t}{t+1} \right) \nonumber
\\
& = & (1-p_2)  \sum_{k=1}^{\infty} g(k)  \left( \frac{1-(1-L)^k}{k L} \right). 
\label{eq:phi}
\end{eqnarray}

Congestion implies that $\phi(L,p_{2})$ is a decreasing function of $L$, so that the more people pass the information on, the less would be the expected value that each of them benefits. The congestion effect gets
  stronger as the number of communicators increases, but  the problem becomes marginally less severe as many informed buyers pass on  the information.\footnote{ Whenever $g(1)<1$, $\phi(L,p_2)$ is decreasing and  convex in $L$,
and each element $ \frac{1-(1-L)^k}{k L}$ is decreasing and convex in $k$. 
Actually, for $k=1$ we have that $ \frac{1-(1-L)^k}{k L}$ is a constant, it is linear for $k=2$, and then it becomes decreasing and strictly convex for each $k \geq 3$.
So, it is strictly convex only if $g(1)+g(2)<1$.
$\phi$ is $1-p_2$ when $L=0$, and then it decreases to $(1-p_2)  \sum_{k=1}^{\infty} \frac{g(k)}{k} $, when $L=1$.}

The two functions $\phi(L)$ and $\Gamma(L)$ are key to understanding the problem of the  seller when setting the bonus  $b$. Indeed, increasing the bonus will induce more people to communicate  the existence of the product to friends, which boosts information diffusion, measured by $\Gamma(L)$, but it also  reduces the expected value of each bonus to the consumers, measured by $\phi (L,p_2)$.
 
Note also that a single informed consumer has no impact on these functions. This will allow us to consider a unique sub-game equilibrium of the game, once the monopolist has chosen prices and incentives.

\section{The trade-off of the seller}\label{tradeoff}

The general problem of the  seller is to maximize the following objective function:\footnote{Marginal cost is normalized to zero without loss of generality.}
\begin{eqnarray}
\label{eq:monopolist_problem1}
\pi(p_1,p_2,b)  =
(1-\beta)  p_1 D_1(p_1) + \beta  \left(p_2-b\right) D_{2}( p_2) \Gamma(L) \ \ , \nonumber
\end{eqnarray}
where $D_{1}(p_{1})$ and $D_{2}( p_2)$ are  the demand of informed and uninformed consumers, respectively.
The first element of this sum is the profit made from informed consumers which, in this setting, is trivially equal to $1-\beta$. Indeed, since the demand of informed consumers is inelastic for all prices below 1 and becomes infinitely elastic above,  the optimal $p_1=1$.\footnote{We refer to \cite{carpinrighi2017} for the study of the model with informed consumers' reservation value distributed as the one of non-informed people. What comes out in this more general case is in favor of the robustness of the simpler model that we analyze here: the optimal $p_1$ for the seller is  the monopoly price that we would get in isolation.} Hence, the problem of the seller can be simplified as:
\begin{eqnarray}
\label{eq:monopolist_problem}
\max_{p_{2},b}\left[
(1-\beta) + \beta  \left(p_2-b\right) D_{2}( p_2) \Gamma(L)\right] \ \ .
\end{eqnarray}
The second element is the profit made on uninformed consumers, which takes into account that,  for each of them reached by information and buying the product, one bonus is paid to an informed consumer. Hence this must be subtracted from  the price $p_{2} $ and acts as a marginal cost in the second term of  equation \eqref{eq:monopolist_problem}. Contrastingly from informed consumers,  an uninformed consumer needs to receive the information (which occurs with probability $\Gamma(L)$) and, once informed, buys only if $v\geq p_{2}$, so that $D_{2}(p_{2})=\Pr(v\geq p_{2})=1-p_{2}$. 

Given the objective of profit maximization, the seller has to choose how to put in place the referral strategy, providing incentives to communication that determine  the fraction of links from informed to uninformed engaging in WOM.
  We assume that the  seller has two  alternatives. The first one is to introduce in the social network incentives to share information about the product through a post on an OSN wall. In this case, once a consumer decides to share the information, this information is  accessible to all people connected with her. For this reason, we label this alternative as \emph{Public WOM}. The second one is to introduce incentives to share information privately friend-by-friend, e.g. by means of instant messaging or emails. We will refer to this second solution as  \emph{Private WOM}.

\subsection{Public WOM}

Under Public WOM, the fraction $L$ is determined by the decisions of each informed consumer. 
An informed consumer who makes a purchase is offered a code or a link to share with friends on her Facebook/Twitter page. Sharing the link on the on-line social network requires the consumer to incur a lump-sum cost $c$ to inform all friends.
 When the  code is used by a friend to buy the product,  the informed consumer receives a bonus $b$.   Given the cost of communication and the bonus, some informed buyers  will pass on the information.
 In particular, an informed consumer with out-degree $k$ will pass on the information only if
\begin{equation} \label{threshold_k}
k  b \phi (L,p_2) \geq c \ \ .
\end{equation}
This means that, for each bonus offered by the seller, there exists a lower level $\underline{k}$ for which informed consumers with that degree are indifferent between passing on or not passing on the information. Given this cutoff,  the fraction of links in which word-of-mouth communication occurs is: 
\begin{equation} \label{threshold_k_L}
L(\underline{k})= \sum_{k=\underline{k}}^{\infty}f(k)
\ \ .
\end{equation}
%
%
Hence, given \eqref{threshold_k_L}, function $\Gamma(L(\underline{k}))$ and $\phi(L(\underline{k}), p_{2})$ depend on  $\underline{k}$. The first function  is decreasing whereas the second one is increasing in the minimal degree required to communicate. Intuitively, if the cutoff $\underline{k}$ drops, the number of active links $L$ increases,  so to  improve  the flow of information - measured by $\Gamma$ - as well as to intensify  competition for successfully passing the information - measured by the decrease in $\phi $. 
Combining  \eqref{threshold_k} with \eqref{threshold_k_L},
$\underline{k}$, in equilibrium, must  satisfy the following equality:\footnote{
Since $\underline{k} \phi (L(\underline{k}),p_2) $ is discontinuous, there are instead some values of $b$ for which no $\underline{k} $ satisfies \eqref{threshold_k} with equality. However, those values for $b$ are clearly dominated choices for the seller.}

\begin{equation} \label{threshold_proof}
\underline{k} \phi (L(\underline{k}),p_2) = \frac{ c}{b} \ \ .
\end{equation}

Notice that  both terms in the left-hand-side of \eqref{threshold_proof} are increasing in $\underline{k}$, so there is a unique $b$ satisfying each value of $\underline{k}$. Hence, the problem of the seller is equivalent to fixing an optimal level of  $\underline{k}^{*}$,  and there is a one-to-one correspondence between $\underline{k}^* \in [k_{f\min},k_{f\max}]$ and the set of optimal $b^*$s. 
So, from  \eqref{threshold_proof}, the problem \eqref{eq:monopolist_problem} of the seller  is equivalent to maximizing:
\begin{equation}
 \label{eq:monopolist_problem3}
\pi(p_2,\underline{k})  = 
(1-\beta)  + \beta  \left(p_2-\frac{c}{\underline{k} \phi (L(\underline{k}),p_2)}\right) (1-p_2) \Gamma(L(\underline{k})) \ \ .
\end{equation}

The solution to the seller's problem is  stated in the following proposition, whose proof (as all the following proofs) is in \ref{app:proofs}.

\begin{proposition}
\label{prop:all1/2}
Every solution to the problem in \eqref{eq:monopolist_problem} is such that $p_2^* =1/2$ and $b = \frac{c}{\underline{k}^{*} \phi (L(k^{*}),p_2)}$ with $\underline{k}^{*}=\arg\max\limits_{\underline{k}}\left[\frac{\Gamma(L(\underline{k}))}{4}-\frac{c\Gamma(L(\underline{k}))}{2\underline{k} \phi (L(\underline{k}),1/2)}\right] $. 
\end{proposition}

Proposition \ref{prop:all1/2} states that, once the  seller identifies the $\underline{k}^{*}$ that maximizes profits, the bonus will be just enough to induce  the buyers with out-degree $\underline{k}^*$ to pass on the information. Once information diffusion takes place, the price offered to newly informed consumers will simply be the monopoly price, i.e., $1/2$.   
The degree distribution is key to determine  the optimal $\underline{k}^{*}$, which trades-off between information diffusion and the size of the bonus. Indeed, lowering the cutoff would induce  less-connected buyers to pass on the information, but would require a higher bonus to incentivize them to communicate.  In the  limit case in which the profit attainable when targeting $\underline{k}^{*}=k_{f\max}$ is non-positive, the WOM communication is never profitable. This occurs when  the bonus required to induce people with the maximal degree   to pass on the information, i.e., $\frac{c}{k_{f\max} \phi (L(k_{f\max}),1/2)}$, is higher than the price $p_{2}$, so to make the seller unwilling to sell the product  to non-informed consumers. In formulas the condition is:
$$\frac{1}{2}<\frac{c}{k_{f\max} \phi (L(k_{f\max}),1/2)} \ \ ,$$
which means that, for given cost of communication,  when the $k_{f\max}$ is sufficiently likely to exist, competition is strong even though minimized. As a consequence,  the bonus required to induce these highly-connected individuals to pass on the information is too high for the referral bonus to be profitable. Therefore, the seller  wants to offer the product to informed consumers only. 
In all other cases, the seller always reaches out   and sells the product to some non-informed individuals.

\subsection{Private WOM}\label{Extensions_cost}

Differently from the case of Public WOM, we consider here the case in which  an informed consumer who makes a purchase is offered a code or a link to share with friends via emails/instant messaging. Sharing the link requires contacting each friend individually. Therefore, differently from Public WOM,   a cost per contact $c_{pr}$ rather than  a lump-sum cost of communication $c$ is faced in order to pass on the information.  Costs per contact  better fit the case of Private WOM rather than Public WOM, where just a click would be sufficient to inform all friends. 
 To capture the idea that contacting a friend individually  requires more effort  and causes more inconvenience than just posting an advertisement on  social media,  it is natural to assume  $c_{pr}  \geq c$.
When  comparing  the two processes, in Section \ref{mix}, we will actually set $c_{pr}= c$.

Similarly to the Public WOM regime, the offer of a bonus induces each agent to pass on the information to a friend if adequately incentivized by $b_{pr}$. The difference is that the decision upon communication here is based on each link rather than on the entire set of links. Namely, an informed buyer will pass on the information to a friend if the expected value of a bonus is sufficient  to offset the cost $c_{pr}$, which is assumed lower than 1$/4$,\footnote{Notice that when the cost is higher,  incentivizing communication is never profitable, as the resulting bonus is always higher than the price.} i.e.:

\begin{equation}\label{real_incentives}
\phi(L,p_{2}) b_{pr}\geq c_{pr}.\end{equation}

Therefore, for any given bonus, the fraction of links from informed  to uninformed consumers in which word-of-mouth communication is used will adjust in such a way that $\phi(L,p_{2}) b_{pr}= c_{pr}$. Indeed, if $\phi(L,p_{2}) b_{pr}$ were lower than $c_{pr}$, informed consumers would face negative utility (excessive communication), whereas if the opposite were true, there would be some links for which communication would  not occur even though it would have made some informed consumers better-off.

Notice that, for the informed consumers, each out-link will be considered as an independent problem, and every interior solution will be actually a mixed equilibrium in which each informed consumer is indifferent between passing on or not passing on the information to each of her links. It is also important  to notice that such mixed equilibrium will be stable, because if we lower $L$, congestion will decrease, and it will be more profitable to pass on the information.
In the same way, if we increase $L$, congestion will increase, and it will be less profitable to pass on the information.

Therefore, if there is a cost per contact, the fraction of links $L$ in which communication occurs  depends on the single local  interaction buyer-friend, so that the degree of the buyer does not play any role.  
In particular, the fraction of links passing on the information, $L$, depends on the incentives of each buyer to pass on the information to \emph{each of her friends}.

The seller anticipates the effect of $b$ on the behavior of informed buyers, so that the problem is equivalent to fix an optimal  $L$, as there is a one-to-one correspondence between $L\in [0,1]$ and $b$, which simply becomes equal to $ \frac{ c_{pr}}{\phi(L,p_{1},p_{2})}$. Therefore, the problem of the seller becomes: 
\begin{eqnarray}
\max\limits_{p_2,L}\pi(p_2,L)  =\max\limits_{p_2,L}\left[
(1-\beta)   + \beta  \left(p_2-\frac{ c_{pr}}{\phi(L,p_{1},p_{2})}\right) (1- p_2) \Gamma(L)\right].
\end{eqnarray}

The equilibrium in the off-line network is expressed in the following proposition.

\begin{proposition}
\label{prop:cost_per_contact}
Consider a cost per contact $c_{pr}$. The seller sets  $p_2^* =1/2$ and $b^{*}_{pr} = \frac{c_{pr}}{ \phi (L^{*},1,1/2)}$, with the unique $L^{*}$ that solves:
$$\frac{\partial \Gamma}{\partial L}\left(\frac{1}{2}-\frac{c_{pr}}{\phi}\right)+\frac{c_{pr} \Gamma}{\phi^{2}}\frac{\partial \phi}{\partial L}=0\\.$$
\end{proposition}

In terms of optimal bonus, the result in Proposition \ref{prop:cost_per_contact} is very similar to the one in Proposition \ref{prop:all1/2}: the seller incentivizes the fraction of buyers that trades-off between profit margins (decreased by $b$) and demand (increased by $b$ through $\Gamma$).

\subsection{Product diffusion: Private vs Public WOM}

In both Private and Public WOM cases, except specific situations, the seller never maximizes product diffusion making every consumer informed. 
This would be the case when $L=1$.
 Intuitively,  maximizing diffusion  would be guaranteed by maximizing  the number of \emph{referrers}. 
%
Maximizing  the information circulating in the network   would not always  be  compatible with the seller making non-negative profits. This depends on the cost of communication and the network density.  To understand how, consider  what would be the share of $b$ that an informed consumer would expect to get from each link, if all informed consumers passed on the information. 
This is 
\begin{equation}\label{exp_rec}
\phi (L=1)=E_k [ 1/k ] /2= \sum_{k=1}^{\infty} g(k)/2k \ \ .
\end{equation} 
$E_k [ 1/k ] $ is a measure of network density, as  the expected reciprocal of the in-degree  is unambiguously  lower as the network becomes denser.

 If the cost of communication is sufficiently small with respect to network density, then the maximal level of information could  circulate on the network with the seller making positive profits.

\begin{corollary}\label{penepene}
If the network is not too dense, it is always possible to inform all consumers making positive profits. 
  Under public communication,  it is needed that $E_k [ 1/k ] >  \frac{4 \beta }{(4-3\beta)}\left(\frac{c}{k_{f \min}}\right)  $.   Under private communication, maximal information diffusion is compatible with positive profits  when $E_k [ 1/k ] >   \frac{4 \beta }{4-3\beta}c_{pr}  $.  
\end{corollary}

\begin{revs} Corollary \ref{penepene} states that,\end{revs}  when  the network is  not too dense in relation to the cost of communication,  it is always possible for the seller to reach the maximal  information diffusion without making negative profits. When this occurs, all those people potentially interested are able to make a purchase. 
However, the network degree distribution is crucial to  align/misalign  information spread and sellers' profits. 
It is possible to establish a simple sufficient condition under which the seller will never maximize the information diffusion about the product, as well as making positive  profits.

\begin{corollary}\label{prop:misalignment}
  Under private communication, the seller will  never maximize the product diffusion  regardless the network density. Under public communication,   if $E_k [ 1/k ] <   \frac{2 c}{k_{f \min}}$  the seller will  not maximize  diffusion. 
\end{corollary}

The intuition is that the seller can maximize both profits and information diffusion only when it is  possible and profitable for her to provide a $b$ that is high enough so that even informed consumers with the lowest possible out-degree will pass on the information. 
However, the bonus that  maximally spreads information is generically non-optimal.   In  particular, the optimal bonus never maximally diffuses  information in the Private WOM nor does it in the Public WOM 
 if the network is sufficiently dense (so that $ E_k [ 1/k ]$ is small).   Hence,  eliciting Public WOM could  maximize  information diffusion as a byproduct of profit maximization  provided that the network is not too dense and therefore congestion is not too severe.   This is never possible under Private WOM. These differences are due to the fact that reaching any level of information is more costly under Private WOM, where the individual incentives to communication are weaker.
 
 These aspects can  be also  noticed by comparing the bonuses and the profits  under the  two communication regimes. 

\begin{remark}\label{Pub_vs_Priv}
Diffusing the same level of information would require a higher bonus under Private WOM. As a result, the profit is always  higher under Public WOM. 
\end{remark}

Remark \ref{Pub_vs_Priv} stems from the fact that any given bonus surely generates more communication publicly than privately, since incentivizing the first type of WOM is obviously cheaper. This is a direct consequence of the intrinsic difference between Private and Public WOM. In the first type of communication,  each consumer has to pass on the information to each contact, which is, for any given degree, more costly than passing on the information through a post on Facebook. As a result, inducing the same level of communication is surely more costly for the seller. Thus, for a given bonus,  the profits that  can be made by eliciting Public WOM are always higher than the ones than can be made under Private WOM. 

\bigskip

To conclude the present section, notice that we  have drawn  attention to the optimal choices of a seller who wants to stimulate either Private or Public WOM,  highlighting  a link between network density and information diffusion. The remainder of the paper  aims to answer  two questions that naturally arise, regarding the network in which  the seller prefers to implement her strategy as well as the strategy itself.

The first question is the following. If either public or private communication are used to spread the word among consumers,   how many channels of information diffusion should be used by a company to run its referral program? Such a decision is ascribable to the one of choosing a denser  or a sparser network. In other words, if the seller has the choice between two different networks, a \emph{sparse} and  a \emph{dense} network, is it possible to move from the former to the latter while increasing profits?  What is the effect of this choice on the information spread? What is the role of hubs/web influencers?   How does this choice differ  between Private and Public WOM? We will provide answers to all these questions in the following section. 
 
 The second  question concerns the mix between Public and Private WOM. 
Remark   \ref{Pub_vs_Priv} seems to suggest that Public WOM is always better for the seller. However, in real markets both types of communication are elicited. For example, referral programs, such as the ones used by Dropbox and Airbnb, give the consumer the opportunity to share the referral link through different media: OSNs, email etc. However, this is usually done by offering a bonus conditioned only on successful referrals and is independent of the channel that generated it. According to our analysis, this seems to be suboptimal, as discriminating bonuses  would take into account the different individual incentives elicited. Namely, the seller could discriminate between public and private communication, offering a bonus whose value depends on whether a given friend is informed through a post  on Facebook or through a private message. This will be discussed in Section \ref{mix}.

\section{Moving to a denser network}\label{virtual}

Companies running a referral program have different alternatives to pursue their objectives. One important  decision is  whether to stimulate WOM in a \emph{denser} or in a \emph{sparser} social network of friendships and acquaintances. 
On the one hand,  \emph{operating in a dense network} improves the information diffusion;   the more people are connected with each other, the easier \begin{revs}it\end{revs}  is to inform  each person about the product.  On the other hand, \emph{operating in a sparser network} reduces competition among buyers, making it cheaper for the seller to incentivize communication. 
The decision of \emph{selling  in a sparser network} can be interpreted in two alternative ways. Under Public WOM, for example, it could be either read as the choice of the seller to stimulate WOM only on the social network provided by Facebook, instead of the combination of Facebook, Twitter, and possibly other social media (which are clearly super-networks compared to just Facebook alone).   Similarly, under Private WOM, it can be interpreted as allowing to pass on the bonus only through emails  rather than emails plus instant messaging.  In general, it can alternatively be interpreted as a limitation of the ego network of each consumer through a  restriction of  the maximal number of bonuses each agent can obtain. 

In the following, we will compare a sparser and a denser social network in order to understand when product diffusion and profits can both be improved moving from a \emph{ sparser network} to a \emph{ denser network}, or when this decision improves on one of the two dimensions but undermines the other one. Moving from the sparser to the denser network simply  means adding links, so that  both in-degree and out-degree distributions $f(k)$ and $g(k)$ receive a shift towards higher density. Because of this, we can formally  use  the concept of first-order stochastic dominance, adopting the definition below provided by \cite{diff2016handbook}.

\begin{definition}
\label{def_fosd}
A distribution $f'$ first-order stochastically dominates (FOSD) a distribution $f$  if,  for every $\hat{k}\in\{1,...,\infty\}$, and for every nondecreasing function $u: \mathbb{R} \rightarrow \mathbb{R}$, it holds that: 
$$\sum\limits_{k=1}^{\hat{k}} u(k) f(k)\geq \sum\limits_{k=1}^{\hat{k}} u(k) f'(k).$$
\end{definition}

 In the following, we will explore the choice within communication regimes, i.e.~Private and Public WOM. In the first case, the seller has to decide whether to move from a sparser to a denser network stimulating Private WOM (Section \ref{mktpene}). We also demonstrate that this situation is qualitatively similar to a case in which the seller focuses only on information diffusion under Public WOM, i.e.,  fixing the degree cutoff for communication  $\underline{k}$.    
Then, in Section \ref{profitmax}, the seller has to decide whether to stimulate Public WOM in a denser or in a sparser network  with the objective of profit maximization. Namely, the cutoff    $\underline{k}$  is allowed to  be adjusted in the denser network to its profit-maximizing level. In this case, the choice of operating in a denser network might be detrimental in terms of information diffusion.

\begin{revs} 
We start by providing Lemmas that will help us to achieve the main results of this section.
\end{revs}
\subsection{Density and Private WOM}\label{mktpene}
We start by analyzing how the choice of going to a denser network affects  profits when the seller stimulates Private WOM. Let us assume a shift in the in-degree distribution to some $g'(k)$ which FOS-dominates $g(k)$. In order to maintain matching between the numbers of  in- and out-degrees as defined in equation \eqref{eq:consistency}, and since we are considering a superset network, also a FOSD shift to some $f'(k)$ is needed. Let us consider the incentives to pass on the information expressed by the inequality in \eqref{real_incentives}. For any given bonus $b_{pr}$ put into the network by the seller, the fraction of active links adjusts until there is no incentive to speak further. Therefore:
$$L=Prob(b_{pr}\phi(L)>c_{pr}) \ \  .$$
For any given $b_{pr}$ and $c_{pr}$, the effect of a FOSD shift from $g(k)$ to $g^{\prime}(k)$ is to increase $\phi(L)$ to $\phi^{\prime}(L)$,  because for any given proportion of active links, each agent has more out-links.  
Therefore, at the original level of $L$ in the sparser network, the equality above no longer holds as $L<Prob(b_{pr}\phi^{\prime}(L)>c_{pr})$. In other words, there is insufficient communication, as some consumers would have incentives to pass on the information and  are not doing so. As a result, for given bonus $b_{pr}$, the level of active links in the new network must adjust to a higher level $L^{\prime}$. As a result, for any given bonus, the seller succeeds in informing more consumers, so that $\Gamma$ increases in the denser network, thus allowing us to conclude the following:

\begin{remark}
Under Private WOM, the seller always makes higher profits in the denser network. 
\end{remark}

\subsection{Density and Public WOM}\label{profitmax}

\begin{revs} The relationship between density and Public WOM is very different from the one discussed in Section \ref{mktpene}, because  the individual perception of congestion is stronger in this type of channel. Indeed, an informed consumer perceives the competition of highly connected peers in Public WOM more than in Private WOM, where decisions are based on single interactions. Therefore, it is not obvious that increasing density under Public WOM would improve neither on diffusion nor on profit. Moving to a denser network implies a possible reduction of incentives to communicate due to congestion even if the bonus remains fixed. Therefore, the seller may find it profitable to limit congestion by staying in a sparse network. Only in the restrictive case in which the threshold for communication does not change with density, it is still possible to conclude that:
\end{revs}


\begin{lemma}
If $\underline{k}^*$ is kept fixed, product diffusion improves  when the nodes are more likely to have higher degrees (as a consequence of a FOSD shift).
\label{FOSDWelfare}
\end{lemma}

\begin{revs} The FOSD shift of Lemma \ref{FOSDWelfare}\end{revs} enhances product diffusion, as the $\underline{k}^*$ does not adjust to its optimal level in the denser network and thus only the information-diffusion effect of the FOSD is at play.  
  Since people are more connected,  more information circulates on the social network and thus more people buy the product. The point is that this would not necessarily benefit profits, as  the impact of higher network density on competition for bonuses is completely disregarded. Any bonus paid to a consumer has the objective of maximizing the set of buyers, without taking into account the competition for bonuses. 

Contrastingly,  the effects of a FOSD increase in $f(k)$ and $g(k)$ on profits are not as unambiguous as they are for information diffusion. This is because, even though  operating in a denser network increases the proportion of activated links $L$, and also the proportion of links receiving the information $\Gamma (L)$,  the  effects on $b=\frac{c}{\phi \underline{k}}$  shift in the opposite direction, as the FOSD  boosts competition between informed consumers. As a result, the seller will be  forced to increase $b$ to compensate for this, if she wants to maintain the same $\underline{k}^{*}$.

 \begin{figure}[h!]
 \centering
       \includegraphics[width=.45\textwidth]{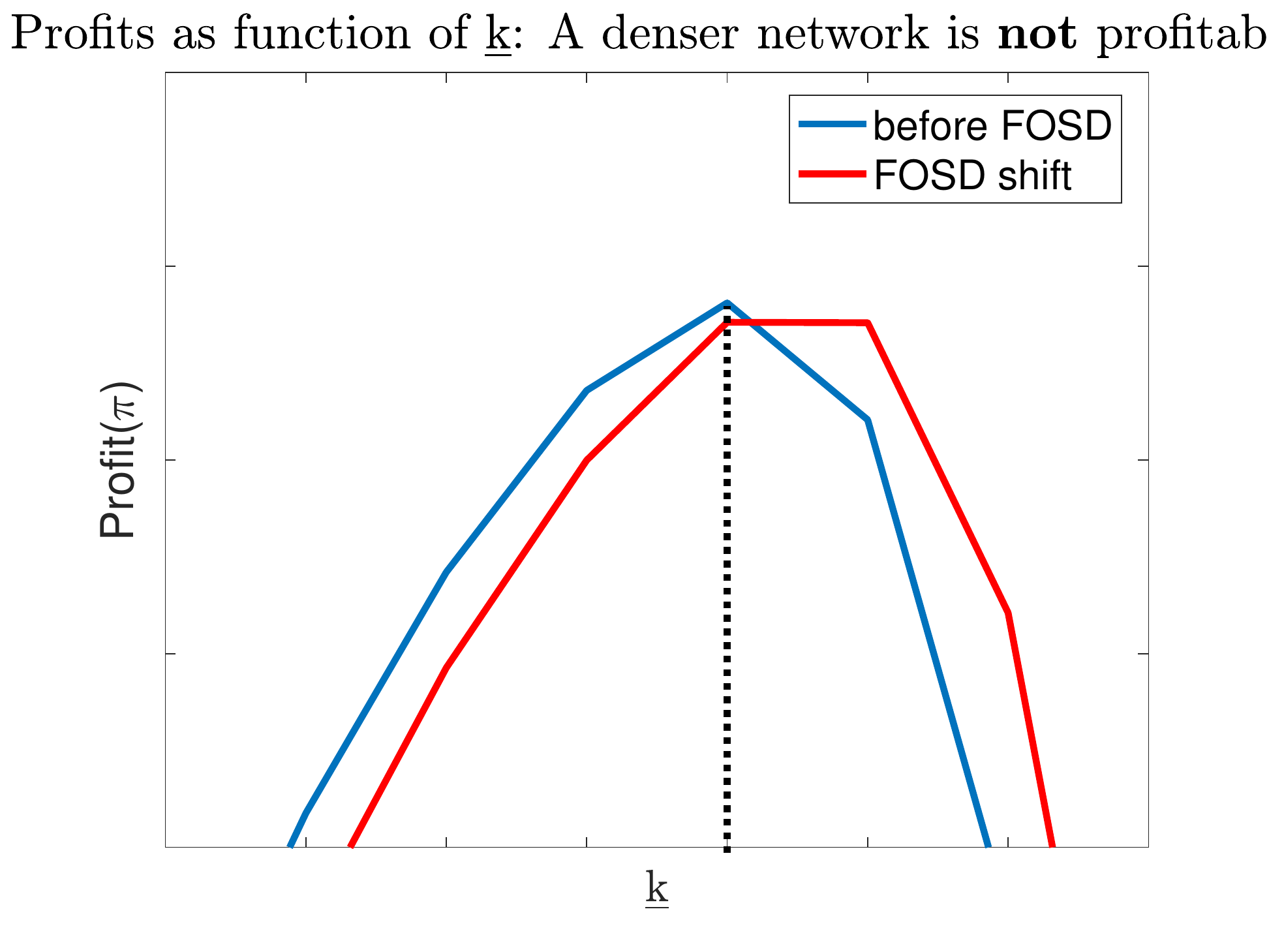}
        \includegraphics[width=.45\textwidth]{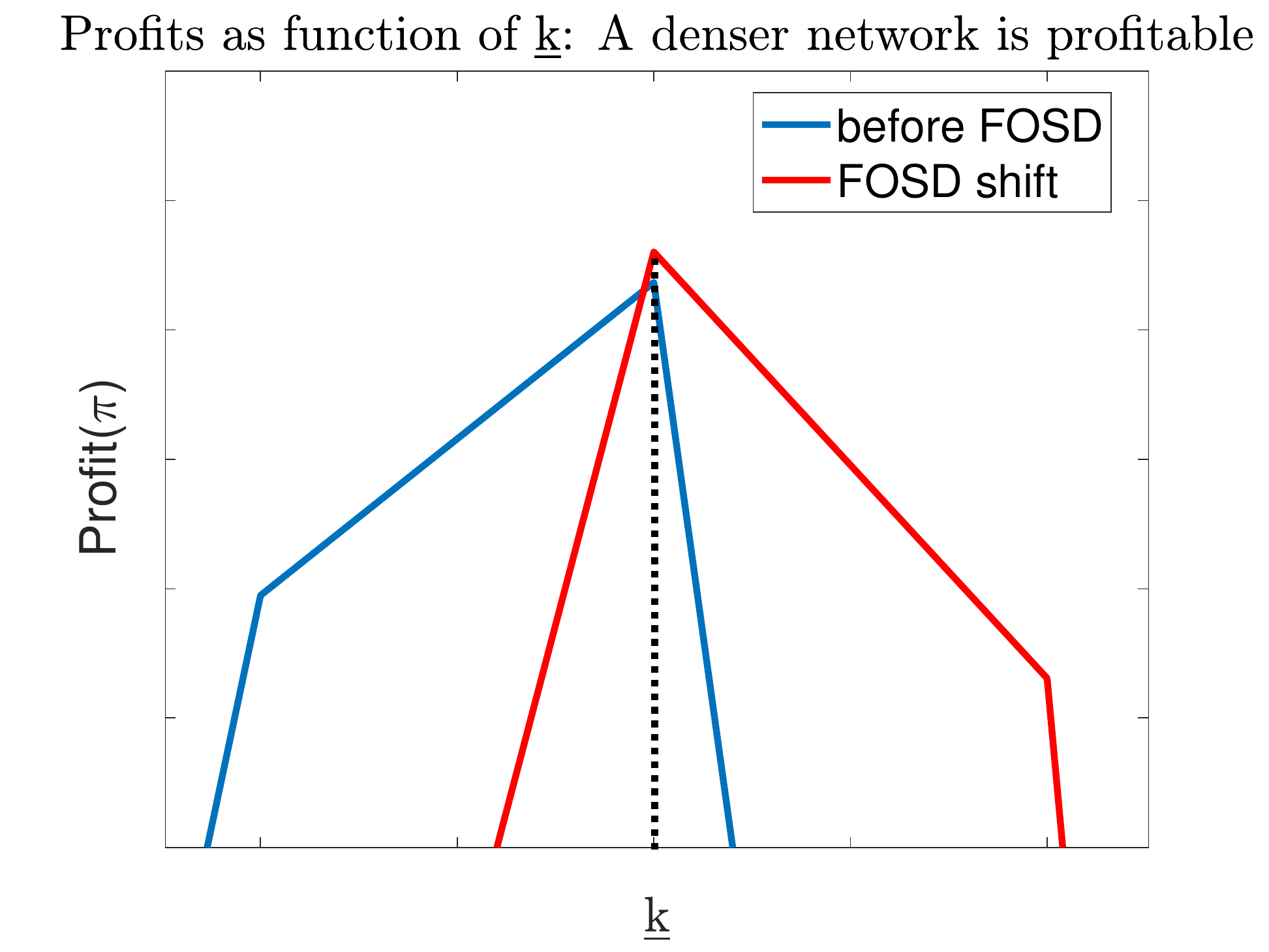}
        \caption{ Left panel: Example in which  operating in the denser network  reduces profits. Both in- and out- degrees are Bernoulli random networks with probability of an in-link $\lambda_{g}=0.1$, and of an out-link  $\lambda_{f}\sim0.0251$ (see \ref{Construction} for details about network construction) and $\beta=0.2$. FOSD is done by increasing $\lambda_{g}$ by $1.6\cdot 10^{-3}$. Right panel: Example in which a denser network is profitable. Both in- and out- degrees are random networks with $\lambda_{g}=0.44$, $\lambda_{f}\sim  0.2931$ and $\beta=0.4$. FOSD is performed  by increasing $\lambda_{g}$ by $1\cdot 10^{-3}$. In both cases network size is 1000. }
\label{examples_profit_going_online}
\end{figure}

In Figure \ref{examples_profit_going_online}, we report  examples of slight FOSD shifts which do not entail any changes in the optimal $\underline{k}$. As it can be observed, there are cases in which moving to the denser network  is profit-enhancing  and others in which it is not. The density of the network and the proportion of uninformed people become crucial, as they determine the gains in terms of information spread that would follow a switch to a denser network.  
The left panel considers a case in which  there are few uninformed people on a relatively sparse network. In this case, a marginal increase in density creates a negligible amount of additional information spread. Moreover, a higher density also strengthens competition among informed buyers, and therefore requires a higher bonus. The balance between competition and diffusion of information makes it not worth it to operate in the denser network. 
Differently, when many uninformed consumers are laid on a denser network (right panel),  information diffusion becomes more salient to the seller. Therefore, she finds it profitable to pay the  higher  unitary bonus needed for passing on the information, as the gains in terms of information diffusion are more significant. 

When setting the optimal $\underline{k}^*$, the seller faces a trade-off between increasing information diffusion (represented by $\Gamma$) and reducing the cost of providing incentives. 
Lowering  $\underline{k}^*$ will boost  $\Gamma$  but will also make competition among informed consumers tougher,  requiring a higher $b$. This comparative-statics analysis is non-trivial and its results strongly depend on the functional forms given to the degree distributions.
The trade-off between information diffusion and competition among consumers is clear-cut  even if not analytically provable in general. Nevertheless, the analysis of  archetypal network topologies helps us to assess the effects of the network density on seller's incentives and optimal decisions. In what follows, we highlight two main aspects. First, we show the emergence of instances  in which operating in a  sparser or a denser network can be better for profits, considering simple cases of degree homogeneity and specific types of FOSD.  Second, we underline the role of hubs in the optimal decision of the seller.
 
\subsubsection{Homogenous degree}

The following lemma considers the simplest case in which all consumers have the same in- and out-degree. This allows us to eliminate any effect of a density increase on the number of consumers who pass on the information.

\begin{lemma}\label{prop:FOSD_homogeneous}
Let the out-degree be $k_f$ and the in-degree be $k_g$ for all agents. 
In this case, moving to a denser network with $k'_f>k_f$ and $k'_g>k_g$ is always weakly profitable for the seller. 
If the seller was using word-of-mouth in the sparser network, then operating in the denser one is always strictly  more   profitable and improves information diffusion.
\end{lemma}

The interpretation of the results of Lemma \ref{prop:FOSD_homogeneous} is straightforward. 
 Indeed, this is a degenerate case in which the absence of variance in the degrees allows the seller to fully extract the surplus of communicators who are all \emph{infra-marginal}. The resulting incentives are clear-cut: the choice of the seller is whether to make people at $k_{f}$ willing to pass on the information for any given network density.  Depending on the cost of communication, this might be profitable or not in the sparser network. As far as moving to a denser network is concerned, there is a positive effect on the information diffusion $\Gamma(L^{*})$ for any given number of targeted senders $L^{*}$, whereas the marginal cost of providing the required incentives remains constant in both networks. Therefore, the seller always  increases profits for any given number of senders. This has to be intended as a weak condition, meaning that the cost of communication might be so high to prevent profitable diffusion of information both in the sparser and in the denser network.   
 However, as discussed in many cases below,  the seller faces a material trade-off when introducing variability in the degrees.

In relation to Lemma \ref{prop:FOSD_homogeneous},  we now relax the assumption of homogeneity in the degree of informed consumers and search for sufficient conditions under which the seller prefers the sparser network. 
First of all, notice that the fraction of senders is required  to not decrease (too much) in relation to the sparser network  for information diffusion to improve  when the seller serves the denser network. Formally: 

\begin{lemma}\label{prop:INHOMOWELF} Consider a situation in which the in-degree distribution is homogeneous. Unless all consumers get informed  in the sparser network, it is always possible to find a sufficiently pronounced FOSD shift  such that operating in the denser networks improves information  diffusion. \end{lemma}

The main important factor for information diffusion is the fraction of people receiving the information, which is an increasing function of $L$. Lemma  \ref{prop:INHOMOWELF} expresses a sufficient condition by focusing on the limit case in which every buyer passes on the information in the original network. If the increase in density 
makes the seller set incentives such that all buyers pass the information also in the denser network, then diffusion  is trivially enhanced, as condition in Lemma  \ref{prop:INHOMOWELF} is always satisfied.\footnote{Notice that when the share of people passing the information in the original network is lower than one, we can also have diffusion-favorable FOSD shifts in which more people are willing to pass the information in the denser network. } 
 On the other side, if less people pass on the information at equilibrium, for the change to inform more people,  it is necessary that the gain in  information diffusion entailed by the higher density overcomes the reduction in the fraction of  communicators. This always happens when the FOSD shift is sufficiently pronounced, so that the condition in \begin{revs}Lemma\end{revs} \ref{prop:INHOMOWELF} is satisfied. However,  there are cases in which the  seller prefers to operate in the sparser network, as we show in  the following Proposition: 
\begin{proposition}\label{prop:HOMOIN}
Consider a situation in which the in-degree distribution is homogeneous.
Consider a FOSD shift such that $k_g$ increases to $k'_g$, and 
$f(k)$ also has a FOSD shift to some $f'(k)$, so that the consistency condition in equation  \eqref{eq:consistency} is maintained.
If $c > \frac{\beta^2}{4(1-\beta) }$, then it is possible to find a distribution $f'(k)$ such that the seller will obtain strictly lower profits in the denser network.
\end{proposition}

In contrast to what was discussed for Lemma \ref{prop:FOSD_homogeneous}, here the absence of variability in the in-degrees is a problem for a seller facing a denser network. Indeed, the congestion is, in a sense, maximal within a given network structure: the non-informed consumers are all connected to the highest number of influencers present with positive probability. Serving a denser network exacerbates the congestion effect.  Thus, if the cost of communication is sufficiently high, we can always find situations in which operating in a denser network  requires such high bonuses  to be profit detrimental compared to those of the sparser network. 
The cost's threshold increases in the proportion of non-informed consumers. As the information problem becomes more severe, the seller is more likely, for any given network, to use bonuses to generate new demand.

\begin{figure}[h!]
 \centering
       \includegraphics[width=.48\textwidth]{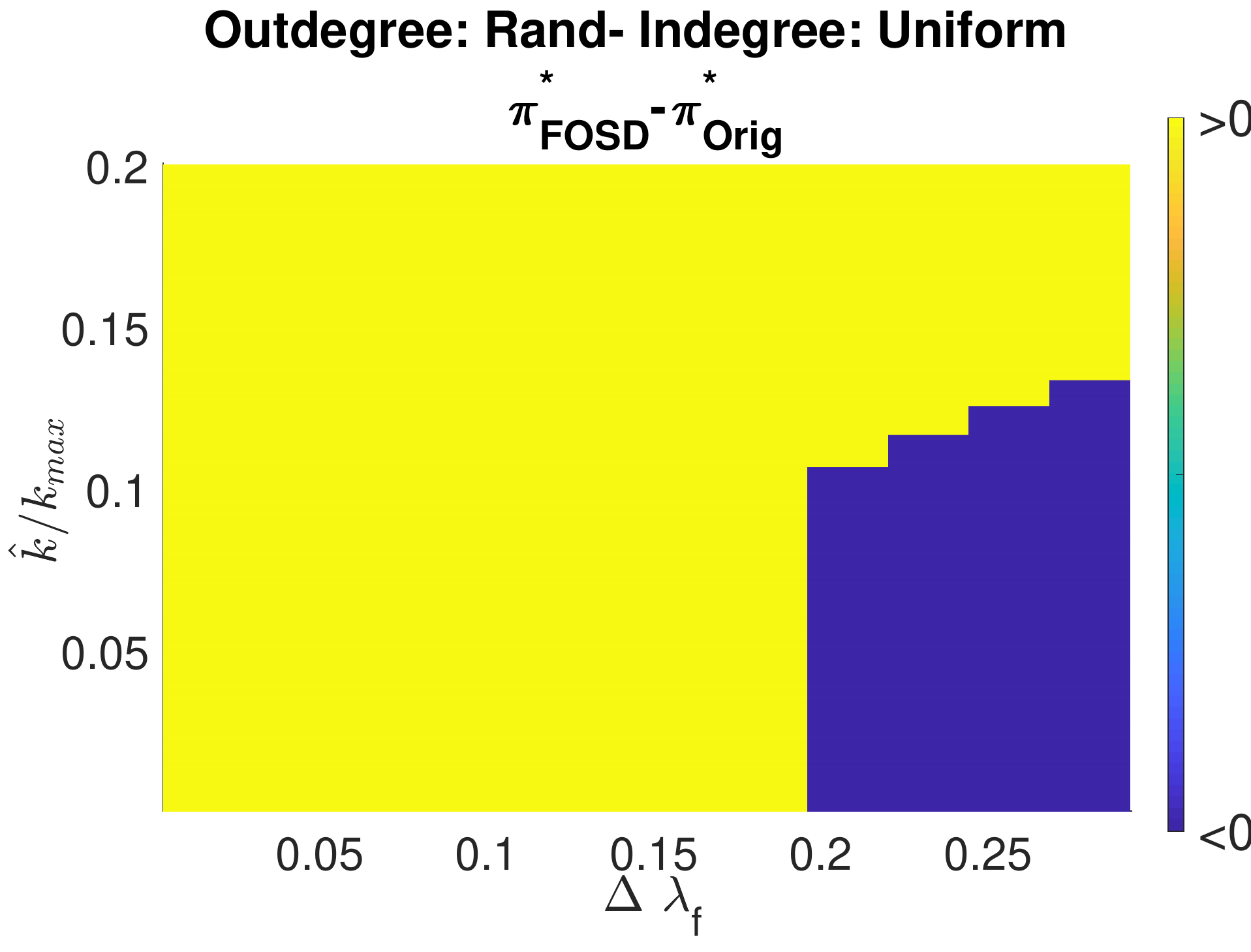}
       \includegraphics[width=.48\textwidth]{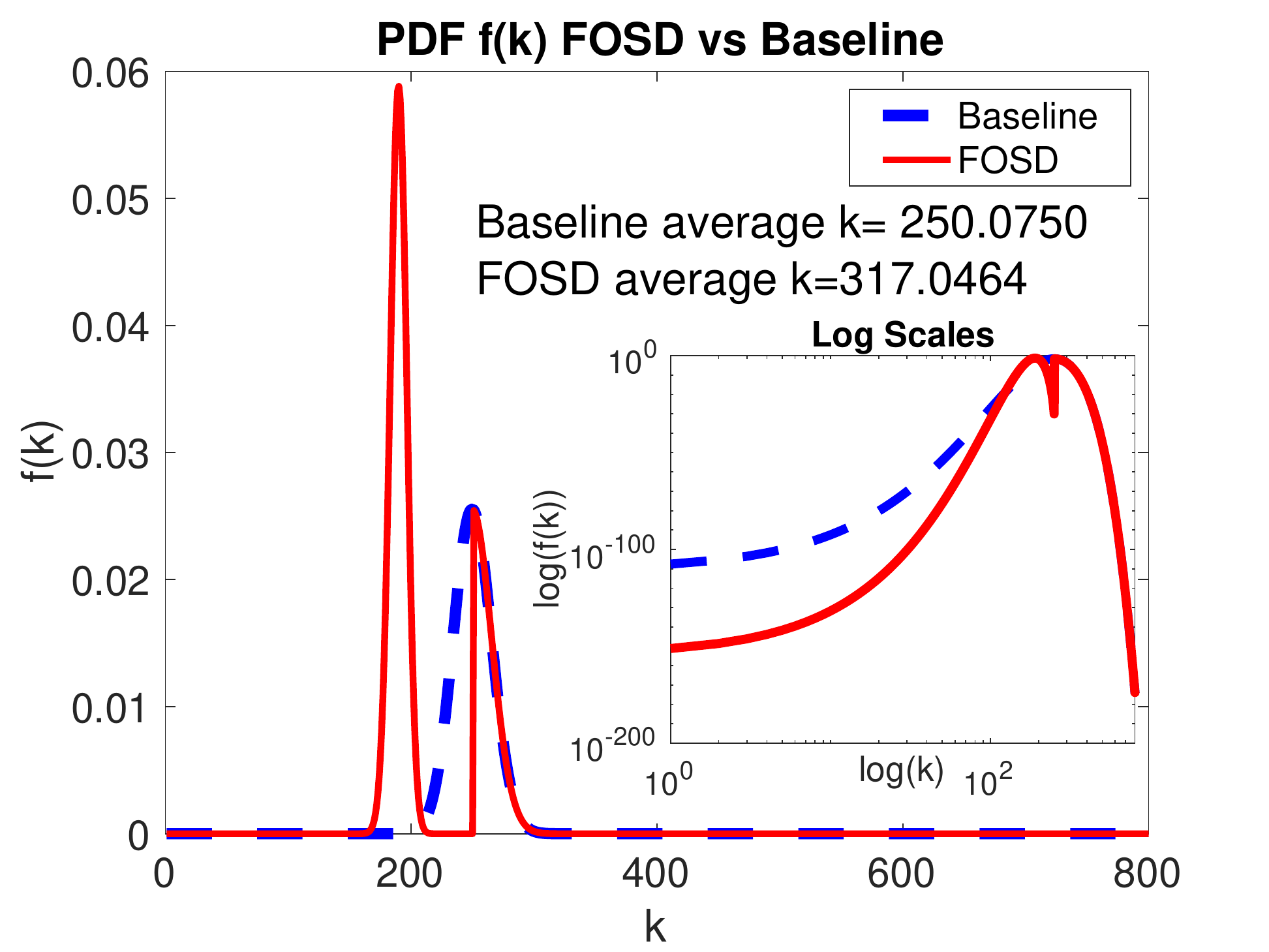}
        \caption{Left panel: The effect of FOSD shift on profits. In-degree is Uniform and out-degrees are distributed as a Bernoulli random network.
We consider FOSD shifts of $\Delta\lambda_f\in[0.01,0.30]$ in steps of $0.025$ (x-axis) for each $\Delta\lambda_f$ we produce a corresponding  FOSD that results by increasing only $f(k)$ for $k\leq\hat{k}$. For every $\Delta\lambda_f$ we study results for $\hat{k}/k_{max}\in[0.0001,0.2]$ in steps of  0.001. In all cases $\beta=0.2$, number of agents: $10000$, $c=0.06$, $k_{max}=n-1$, $k_{min}=1$, $\lambda_f(baseline)=0.10$. Right panel: PDF for an example FOSD with $\Delta\lambda_f=0.10$ and $\hat{k}/k_{max}=0.025$, the initial $\lambda_f(baseline)=0.10$. Inset: log-log representation of the  PDFs.
Yellow indicates   the seller's  preference for  the denser network, blue indicates   the seller's  preference for  the sparser network.}
        \label{strange_fosd}
\end{figure}

\begin{revs} To exemplify  Proposition \ref{prop:HOMOIN}, we numerically study a network in which the in-degree is homogeneous and we assume the out-degree to follow a Bernoulli distribution with probability of a link to exist equal $\lambda_f$. \end{revs}
In Figure \ref{strange_fosd}, we consider a class of FOSD shifts such that the in-degrees increases uniformly, while all additional out-links are added only to nodes relatively poorly connected, that is with degree lower than $\hat{k}$.\footnote{All our numerical explorations henceforth are based on stylized versions  of empirically observed social networks, i.e., Bernoulli random and scale-free network. In Bernoulli random networks the density parameter, i.e. the probability of each link to exist, is named $\lambda$. In scale-free networks the slope of the power-law describing the distribution is  named $\gamma$. Since we work with directed networks, these two classes of distribution are used to model in and out degrees, and we refer to them by using subscripts $g$ and $f$, respectively. \ref{Construction} specifies the functional forms of the network distribution considered.
The Matlab code of the simulations can be found at: \url{https://github.com/simonerighi/CarroniPinRighi_ManagementScience2018}.} 
In practice, it is as if a channel is included in the network that provides more communication to only those informed consumers that had originally a lower out-degree.
What we obtain in this case is reported in the claim below.

 \begin{claim}\label{Claim_bad_FOSD}
Consider homogeneous in-degree and out-degree distributed according to a binomial distribution (Bernoulli random network).  Consider a FOSD shift such that $k_g$ increases to $k'_g$. Accordingly, add all the additional out-links below an out-degree $\hat{k}$, keeping $f(k)$ constant above $\hat{k}$. Then, if $\hat{k}$ is sufficiently low and the FOSD shift is sufficiently pronounced, the seller prefers to operate in the sparse network. 
\end{claim}

Claim \ref{Claim_bad_FOSD} says  that if the denser network is a  FOSD shift as the one  described in Figure \ref{strange_fosd},   operating in a sparser network would make the seller better-off. 
 This FOSD exacerbates congestion among informed people without adding more in terms of information diffusion. As a consequence, the sparse network is preferred by the seller.

\subsubsection{The role of hubs}\label{HUBS}
It is well known that hubs can be powerful drivers of information diffusion due to their disproportionate large degree (see, for example, \citealt{AceBimp}).  In this section, we analyze the role of hubs in \begin{revs} relation with \end{revs} the preference of the seller regarding network density and with 
\begin{revs} the associated product diffusion. The hubs have two important effects on information diffusion in our setup. The first is that they provide the seller with the opportunity to reach a large number of uninformed consumers at a very low cost. This is due to the fact that, when they are the only ones motivated to engage in WOM, they do not suffer congestion. The second is that they are the main source of congestion due to their large number of links. Thus, in presence of hubs, the latter problem becomes of second-order importance when moving to denser networks.

To better highlight the consequences of the first effect, consider a \end{revs} FOSD shift characterized by the introduction of highly connected influencers. We are able to show that it is possible to have denser networks associated with a decline in information diffusion:
\begin{proposition}\label{prop:general}
Suppose that $k_{f \max}$ is finite, and that the seller chooses $\underline{k}^{*}$ such that $\Gamma(L(\underline{k^{*}}))=1-\delta$, with $\delta  <  \frac{2 c}{ \underline{k}^*\phi \left(\underline{k}^{*}\right)}$.
In this case, it is always possible to consider a FOSD shift of the network, such that the seller maximizes profits in the denser network, but diffusion  is lowered by this choice.
\end{proposition}

As formally explained in the proof of Proposition \ref{prop:general}, an example of FOSD shift causing the described scenario is when we add  a few very influencing people (super-hubs or web-influencers) to the original network, redistributing the additional in-links homogeneously.
If the seller incentivizes only these super-hubs to diffuse information, the latter would suffer very limited congestion, making their required bonus cheap. \begin{revs}This bonus-saving strategy may result in lower diffusion: reducing congestion becomes of first-order importance with respect to diffusing information.

 Oppositely, if hubs are present and sufficiently important already in the original network,  they are the predominant source of congestion. Therefore, an increase in density makes congestion of second-order importance with respect to product diffusion\end{revs}. 
In what follows, we \begin{revs} elaborate on   this by providing a  numerical comparison of  the effect of density in different classes of in- and out-degree distributions. \end{revs}
Overall, the numerical exploration of the model supports the following claim.
\begin{claim}\label{claim2}
Fixing a FOSD shift,  moving to the denser network is  profit-enhancing except in very specific cases. It becomes also effective in information diffusion as we move from a network without hubs to a network with hubs. 
\end{claim}

 \begin{figure}[h!]
 \centering
       \includegraphics[width=.48\textwidth]{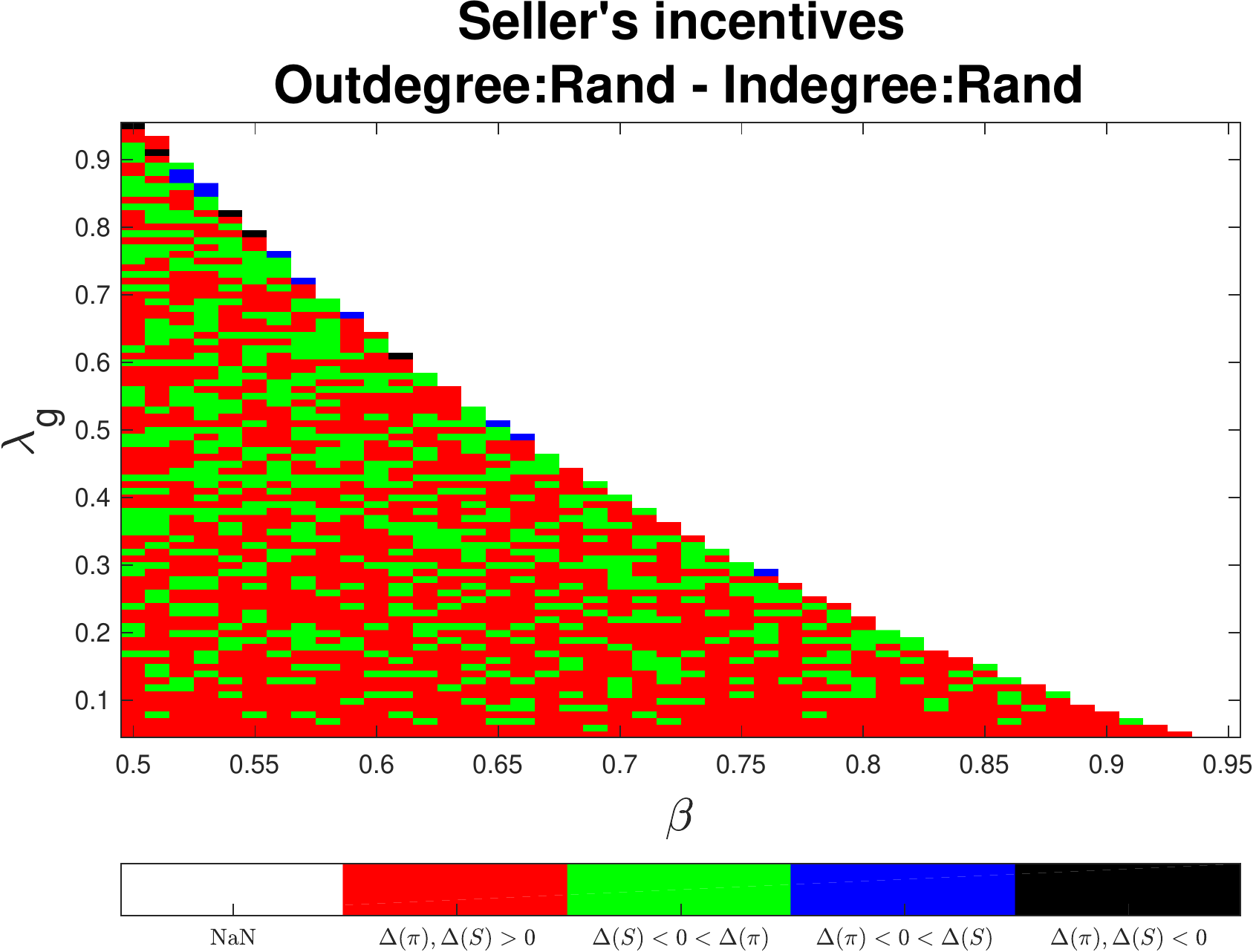}
        \includegraphics[width=.48\textwidth]{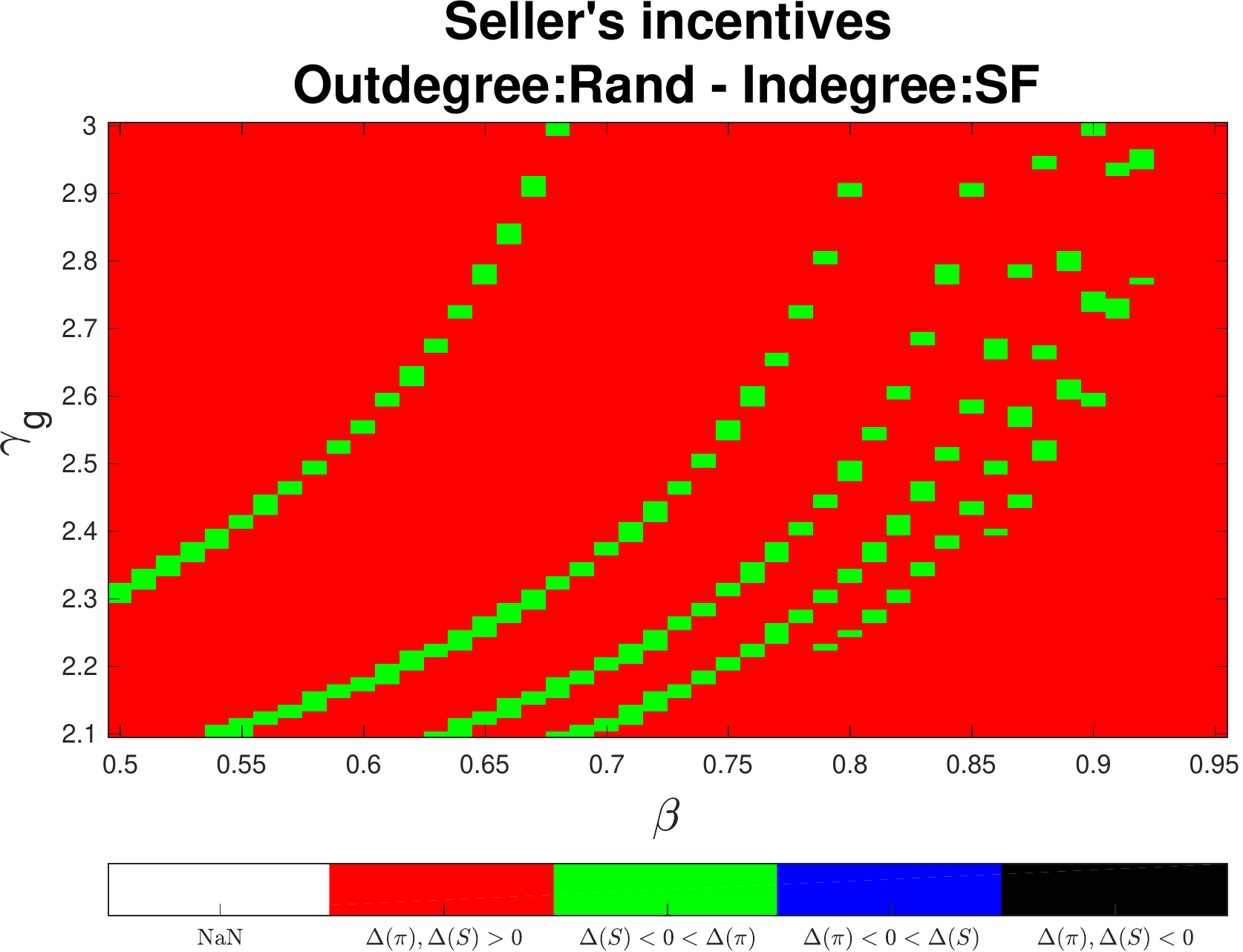}\\
         \includegraphics[width=.48\textwidth]{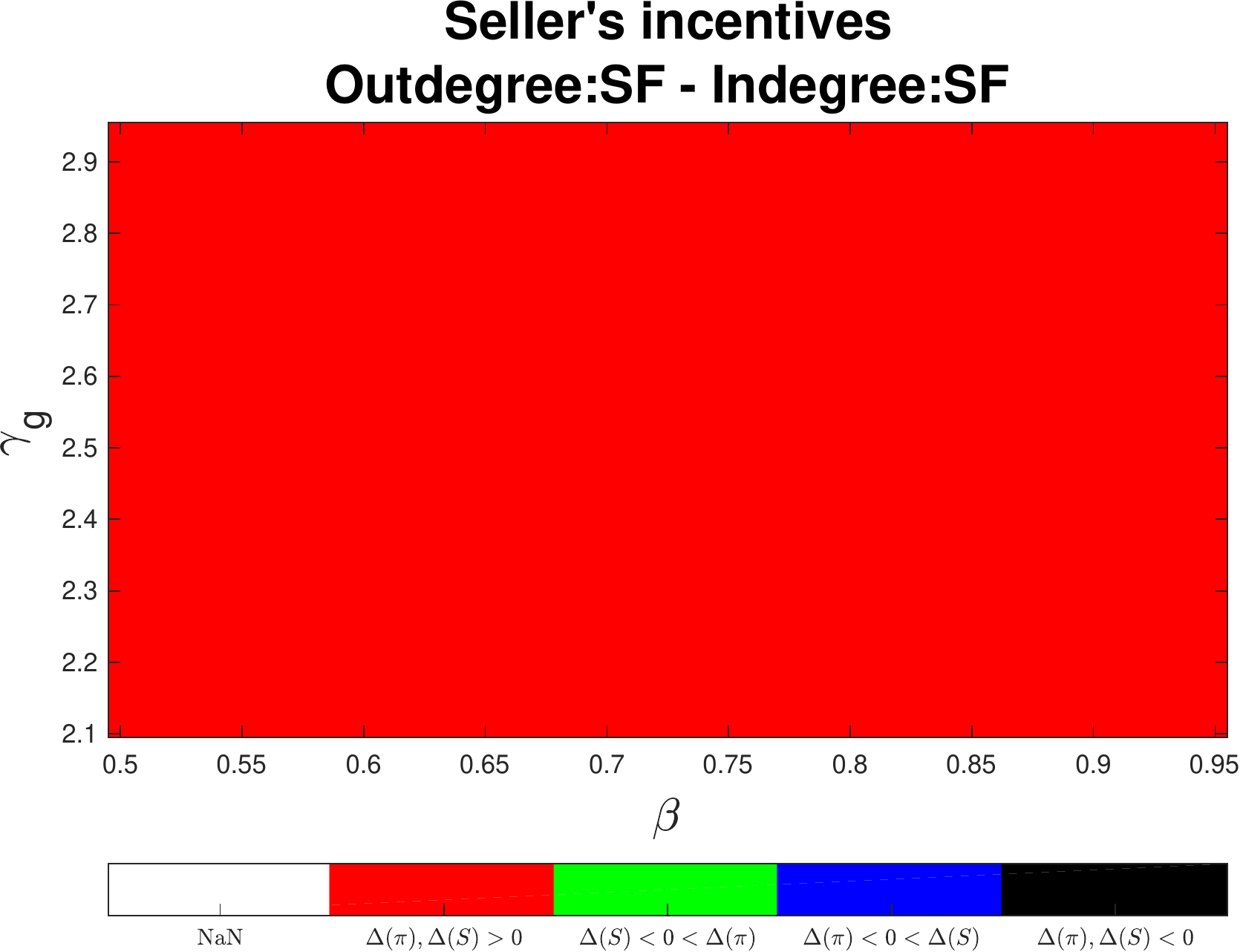}
        \caption{Effect of FOSD shift on profits and on the share of informed consumers who buy the product (S). Left panel: In- and out-degree distributions  are random. Right panel: Out-degree distribution  is random while the in-degree one  is scale free. Bottom panel: In- and out-degree distributions  are scale free.
For random in-degrees, we consider   FOSD shift of $2.5\cdot 10^{-2}$ for all combinations of starting $\lambda_g\in[0.05,0.95]$ in steps of $0.01$. For scale-free in-degrees,  FOSD shift of $2.5\cdot 10^{-2}$ for all combinations of starting $\gamma_g\in[2.1,3]$ in steps of $0.01$. In both cases $\beta\in[0.5,0.95]$ in steps of $0.01$.  Number of agents: $10000$, $c=0.06$, $k_{max}=n-1$, $k_{min}=1$.  }
        \label{sc_fosd}
\end{figure}

As shown in Figure \ref{sc_fosd} (left panel), \begin{revs} when there are no hubs, \end{revs}serving the denser network  is surely profitable for the seller as long as  the original distribution of in-links is not too dense.\footnote{Notice that when in-links are very dense operating in a denser network  becomes less likely to be attractive than operating in a sparser one (blue and black areas in left panel). }
\begin{revs}When in the denser network, the seller anticipates the stronger congestion, and this may make the  optimal targeted $\underline{k}$ shifts upwards in order to partially offset the increase in congestion (green areas). This may result in less information circulating in the denser network, with detrimental  consequences on product diffusion. Hence the absence of a discernible pattern in Figure \ref{sc_fosd} (left panel). \end{revs}

Considering instead cases where the in-degree distribution presents hubs (\begin{revs}right and lower \end{revs} panels of Figure \ref{sc_fosd}).
\begin{revs} In the originally \emph{sparse} network, few hubs attract most of the inflow of information: these people are the main source of congestion for influencers. Once a denser network is faced, the increase in congestion is smaller in comparison to the previous case, as these ``monopolizers'' of information still remain the main source of congestion.  Therefore, diffusion tends to become of first-order importance for the seller. This explains the disappearance of most green areas.  
Finally, when hubs are present also in the out-degree distribution (Figure \ref{sc_fosd}, lower panel), the objective of diffusion fully overshadows congestion effects and  density improves both profits and diffusion. \end{revs}

\bigskip

To sum up, and compare the theoretical results from the previous section with the outcome of the numerical explanation, the following can be said about Public WOM.
 
When both influencers and influenced networks have homogeneous degree distributions, profit maximization is guaranteed by operating in a denser network and this also enhances market penetration (Lemma \ref{prop:FOSD_homogeneous} and Claim \ref{Claim_bad_FOSD}). This is because in the denser network, the demand is higher without increasing the cost of word-of-mouth stimulation.  Differently, when the network of influencers is not homogeneous, it is always possible to find situations in which, if the cost of communication is sufficiently high, higher network density induces too much congestion (Proposition \ref{prop:HOMOIN}). 
Moreover, in presence of online super-hubs,  the optimal profit-maximizing strategy would be to diffuse information only through web-influencers, possibly at the expenses of reducing diffusion of information (Proposition \ref{prop:general}).  The crucial role of hubs is confirmed through an extensive analysis of numerical solutions on a wide class of networks (Claim \ref{claim2}).   When hubs are not important in either networks, any outcome can potentially emerge, i.e., we can have situations in which a denser/sparser network  is optimal for  profits/diffusion, depending on parameters. The picture changes when the sparse network is characterized by few hubs \begin{revs}. In this case, congestion becomes less important so that  the seller mainly focuses on diffusion.\end{revs}

\section{The optimal mix between Public and Private WOM}\label{mix}

So far, we have considered the stimulation of  Public and Private WOM separately.   \begin{revs}  However, in the real world, Public and Private WOM are always mixed. This is due to the fact that   many sellers use both simultaneously and, regardless of seller's choices,  individuals are always  able to pass on the information both publicly and privately. However, Private and Public WOM are important benchmarks to look at because they highlight the fact that engaging in either type of communication implies different incentives. This has important consequences in terms of strategy, as there is scope for discriminating rewards depending on the type of  WOM a consumer engages in. Namely,  Private WOM  is more costly than Public WOM from the consumer viewpoint. Therefore, one would expect that a seller could offer a high reward in the private channels,  so that people relatively more popular  pass the information publicly, and Private WOM becomes  a residual means of information diffusion. 
\end{revs}

For the sake of exposition, let us modify the profit  of the seller as follows:\footnote{$p_{1}^{*}=1$ as in Public and Private WOM studied separately.} 
\begin{equation*}
 \label{eq:prob_{mix}}
\pi(p_2,b_{pub},b_{priv})  = 
(1-\beta)  + \beta  \left(p_2-\mathbb{E}[B|b_{pub},b_{priv}]\right) (1-p_2) \Gamma(L) \ \ .
\end{equation*}
where $\mathbb{E}[B|b_{pub},b_{priv}]$ is the expected amount paid by the seller for each successful  referral. With respect to the previous cases, two main differences emerge. Firstly, the fraction of links from informed to uninformed consumers $L$ is composed by links activated both privately ($L_{priv}$) and publicly ($L_{pub}$). Secondly, given the individual incentives to pass on the information  privately and publicly, the seller may end up in a situation in which a private bonus $b_{priv}$ or a public bonus $b_{pub}$ have to be paid to the referrer.  In order to pin down these values, we  assume that an informed  consumer faces the same cost to pass on the information  publicly (to all her friends) or privately (to one of her friends), i.e.,  $c_{pr}=c$.\footnote{ Notice that this assumption is made in order to study the most extreme case, while all results would hold \emph{a fortiori} assuming $c_{pr}>c$.} 

Let us first analyze the case of Public WOM. 
An agent with degree $k$ has to decide whether to pass on the information publicly or not.   When the code is used by a friend to buy the product,  the informed consumer receives the bonus $b_{pub}$. This will happen with a probability $\phi (L_{priv}, L_{pub},p_{2})$, that this time depends not only on the number of other informed consumers passing on the information publicly but also privately.

    Given the cost of communication and the bonus, some informed buyers  will pass on the information.
 In particular, an informed consumer with out-degree $k$ will pass the information only if
\begin{equation} \label{threshold_k_mix}
k  b_{pub} \phi (L_{priv},L_{pub},p_2) \geq c \ \ .
\end{equation}
Similar to the case of Public WOM only,  each bonus offered by the seller maps into a  cut-off of minimal degree  $\underline{k}$  needed to be willing to pass on the information, so  that  the fraction of links in which Public WOM occurs is: 
\begin{equation} \label{threshold_k_L_mix}
L_{pub}(\underline{k})= \sum_{k=\underline{k}}^{\infty}f(k)
\ \ .
\end{equation}
with $\underline{k} b_{pub}  =\frac{c}{ \phi (L(L_{priv},L_{pub}),p_2) } $ given that,  by equation  \eqref{threshold_k_mix}, any bonus which does not make  equation  \eqref{threshold_k_mix} hold with equality would leave an extra surplus to the buyer.

On the private channel, the fraction of links in which Private  WOM occurs is the $L_{priv}$ such that, for any given bonus, no additional link is worth being activated, i.e., 

\begin{equation} \label{priv_mix}
b_{priv}\phi (L(L_{priv},L_{pub}),p_2) =c
\end{equation}

Notice that equations  \eqref{threshold_k_mix} and \eqref{priv_mix} say that 
 for any $b_{priv}$ and $b_{pub}$ the cutoff $\underline{k}$ will adjust in such a way that $b_{priv}=b_{pub}\underline{k}$. If the private bonus increases (so that also the ratio $b_{priv}/b_{pub}$ increases),  people are more likely to be  willing to pass on the information privately. This increases competition for bonuses so that the expected value  $\phi$ is lower. Thus, if the bonus $b_{pub}$ is kept fixed, the incentives to pass the information on publicly are lower, so that $\underline{k}$ goes up. 
 
 \begin{revs}
 This also implies that equations  \eqref{threshold_k_mix} and \eqref{priv_mix} essentially express the same incentives for informed consumers.  The difference between Public  and Private  WOM is that while the individual characteristics (degree) is key to share a link in the first, the second one only depends on the size  of the reward and on the cost of passing on the information. For this reason, it could well be that the same individual shares the link both publicly (because sufficiently popular) and privately (because the expected benefit outweighs the per-contact cost). In aggregate terms, the fraction of links passing on the information publicly and privately will be such that $L_{pub}=L_{priv}$, since incentives to WOM are precisely the same.
 \end{revs} Therefore, exploiting equation  
 \eqref{threshold_k_L_mix}, we can pin down   the expected  fraction  of links passing on  the information overall, i.e.: 
 
$$L(\underline{k})=\left(\sum_{k=\underline{k}}^{\infty}f(k)\right)\left(1+\sum^{\underline{k}}_{k=k\min}f(k)\right).$$
Notice that the first element of the sum above  is the proportion of links in which public information exchange occurs, whereas the second element is the expected fraction of links in which only private communication occurs. Indeed, since Private WOM depends on the single interaction buyer-friend, the degree of the buyer does not play any role. Therefore, a buyer of any degree is equally likely to pass on the information privately in each interaction. As a consequence, the probability of people passing the information only privately is $L_{pub}$ multiplied by the proportion of links not passing the information publicly.  

To break ties, we assume that if an agent passes on the information to a friend both publicly and privately, the probability that the seller pays bonus $b_{pub}$ or $b_{priv}$ is equal to $1/2$, depending on the code used by the friend.\footnote{One may also assume a strategic behavior of consumers, who then would always use the link leading to the higher bonus for their friend. In that case, the probability of paying $b_{priv}$ becomes 1. This would not introduce any qualitative change in our results.} Accordingly, the number of bonuses $b_{pub}$ expected to be paid at equilibrium is given by the fraction of links in which public communication occurs multiplied by $1/2$, to take into account that the remaining $1/2$ of successful referrals would require a payment of $b_{priv}$. Differently,  the fraction  of links in which communication is only private, i.e., $\left(\sum_{k=\underline{k}}^{\infty}f(k)\right)\left(\sum^{\underline{k}}_{k=k\min}f(k)\right)$, will surely require the seller to pay a bonus equal to $b_{priv}$. As a result, given that $b_{pub}=\frac{c}{\phi\underline{k}}=\frac{b_{priv}}{\underline{k}}$, the expected amount paid by the seller in case of successful referrals is: 

\begin{equation*}\label{expected_bonus}\mathbb{E}[B|\underline{k}]=\frac{c}{\phi}\left(\sum_{k=\underline{k}}^{\infty}f(k)\right)\left[\frac{1+\underline{k}}{2\underline{k}}+\sum\limits_{k=k\min}^{\underline{k}}f(k)\right]
\end{equation*}

Hence, the problem of the seller is equivalent to: 
\begin{equation} 
 \label{eq:monopolist_problem_mix}
\max\limits_{p_{2},\underline{k}}\pi(p_2,\underline{k})  = 
\max\limits_{p_{2},\underline{k}}\left[(1-\beta)  + \beta(1-p_{2})
(p_{2}-\mathbb{E}[B|\underline{k}])\Gamma(L(\underline{k}))\right],
\end{equation}
whose  solution   is  stated in the following proposition:

\begin{proposition}
\label{prop:allmix}
If the seller discriminates between Private and Public WOM, then the optimal price is the monopoly price
 $p_2^* =1/2$, the public bonus is $b^{*}_{pub} = \frac{c}{\underline{k}^{*} \phi (L(\underline{k}^{*}),p_2)}$  and  the private bonus is $b_{priv}^{*}=\frac{c}{ \phi (L(\underline{k}^{*}),p_2)}$
  with $$\underline{k}^{*}=\arg\max\limits_{\underline{k}}\left[\left(\frac{1}{2}-\mathbb{E}[B|\underline{k}]\right) \frac{\Gamma(L(\underline{k}))}{2}\right]
.$$ 
\end{proposition}
Proposition \ref{prop:allmix} shows the optimal discriminating bonuses, which are generically of different sizes. Indeed, the optimal $\underline{k}^{*}$ has to be read  as the ratio between the bonus offered for successful referrals coming from private versus public communication. \begin{revs} In terms of companies' strategies, setting bonuses  in such a way that $\underline{k}^{*}=1$ would mean to adopt a non-discriminating strategy, whereas the extent of discrimination  becomes increasingly  relevant as $\underline{k}^{*}$ goes up.
 Since the $\underline{k}^{*}$ actually  depends on the structure of the social network considered, it is worth understanding what would be  the impact  on the optimal  discriminating strategy of a change in  network characteristics.
  
In particular, the presence of extremely connected influencing hubs  (or web influencers,  using  the jargon of  today's digital markets) together with network density are  the keys to understand how much optimal rewards should differ between channels. In order to focus on the role of these extremely connected individuals, let us set 
 the out-degree distribution as a scale-free and the in-degree as a Bernoulli.

 \begin{figure}[h!]
 \centering
       \includegraphics[width=.48\textwidth]{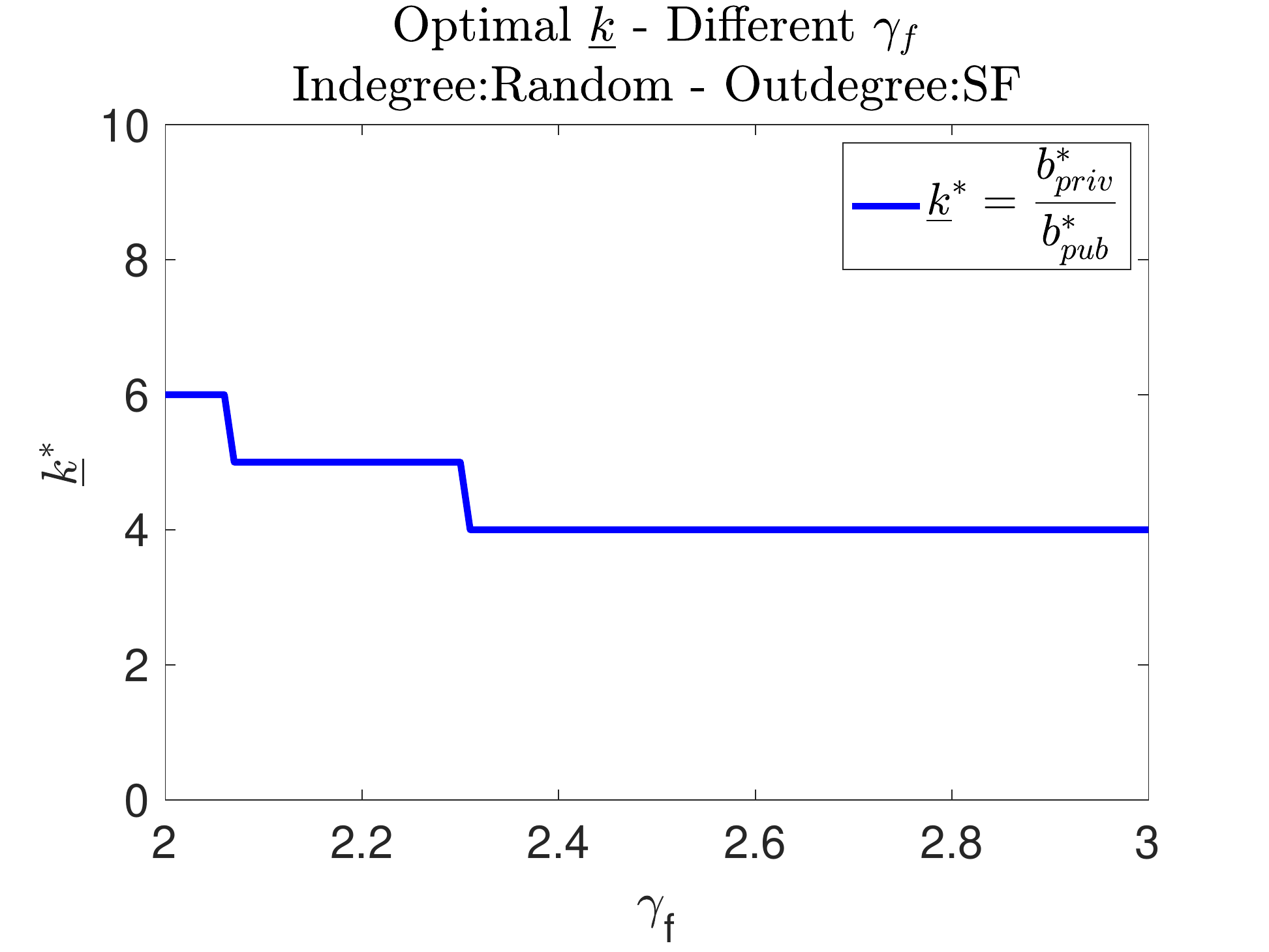}

        \caption{Optimal ratio between bonuses  $\frac{b_{priv}^{*}}{b_{pub}^{*}}=\underline{k}^*$ for all $\gamma_{f}$. Out-degree distribution is scale free while the in-degree one  is Bernoulli random. Effect of different $\gamma_f\in[2.01,3.5]$ in steps of $0.01$. $n=10000$, $k_{fmin}=1$, $k_{fmax}=100$.
}
        \label{mixedstrategy}
\end{figure}

Figure \ref{mixedstrategy} depicts the optimal $\underline{k}^{*}$   for all possible slopes $\gamma_{f}$ of the power law out-degree distribution.\footnote{We refer the reader to   \ref{Construction} for all details about classes of complex networks.}  Going from left to right, we progressively increase $\gamma_f$, thus making hubs more important in the distribution and reducing the density of the network (\citealt[Ch. 4.6]{barabasi2016network}).
Figure \ref{mixedstrategy}  shows an inverse relationship  between the importance of
hubs and the  extent of discrimination, which is due to two reasons. On the one hand, the reduced density requires stronger incentives in order to improve product diffusion on the public channel, which is the most important channel of information diffusion.
 On the other  hand, the fact that hubs become important reduces the communication   incentives of  poorly connected individuals, who suffer the congestion of hubs. Both forces go towards the direction of increasing the bonus offered in the public channel, so to reduce the  difference in optimal  bonuses between  channels.  \end{revs}

\section{Conclusion}\label{conclusions}

Network-related referral strategies are an important market-penetration tool for today's companies.
The use of these strategies in many different markets has exponentially grown thanks to online platforms and social media. Indeed, the latter has enhanced the outreach efforts of companies, giving them an effective and relatively inexpensive way to inform consumers about the products they sell.  Referral strategies seem to be win-win solutions when a seller faces a partially uninformed pool of consumers  as the former increases its demand incentivizing informed consumers with rewards, while uninformed people are made aware about possibly valuable products.


Many potential channels are available for companies interested in introducing referral programs. OSNs, SMSs, emails and instant messaging  are all means that, together, can help a  firm become known to consumers in a very effective way. Nevertheless, companies often do not exploit all these possible channels, but rather limit their strategy to only some of them or, alternatively, cap rewards as to restrict  the ego-network of consumers. At first sight, a more penetrative strategy, maximizing information diffusion by fully exploiting all informed consumers' links seems to be trivially profit maximizing, especially since most firms condition influencers' rewards based on the purchases of influenced people.

In this paper, we show that the optimal strategy of the firm is not trivial and it is always a joint combination of two channels of diffusion.
One is such that the cost of sharing the information is fixed for the informed consumer, as is the case for online social networks.
The other is such that the cost is proportional to the actual peers that the informed consumer contacts, as is the case of face-to-face interaction and SMS messages. Thus, the second type of communication is not influenced by consumer's network position, while the first entails a strong impact of the network on decisions.\footnote{\begin{revs}
One of our results is that the most connected individuals are the key individuals to be motivated to maximize profits. This follows from the fact that we assume a limited knowledge of the social network. When firms know the position of individuals in the relevant social networks, more complex types of centrality (e.g.,  Bonacich centrality as in  \citealt{ballester2006s}, or variants of betweenness centrality  as in \citealt{banerjee2013diffusion}) can be used to improve outcomes. This however can be costly or incomputable  to do in practice, providing a rationale for  companies that use random seeding \citep{akbarpourjust}. \end{revs}
} 
As a consequence, the firm may not want to address a network that is too  dense (e.g., include the possibility of sharing the news with Facebook {\bf and} Twitter), but may prefer a sparser network (e.g., only for Facebook).

The rationale for firms to limit the scope of potential activation of influencers' ties (or to simply stay in a \emph{sparser} network) is congestion, i.e., the fact that, in a network, non-informed people may receive information by multiple sources. As a result, the more numerous the buyers who become active influencers, the harder it is -  for each of them - to obtain a bonus. Therefore, a seller has to consider the trade-off between demand size (information spread)  and the marginal cost of attracting that demand (giving the right incentives through bonuses). In this regard, network density is crucial and has two opposite effects. On the one hand,  it makes congestion stronger,  thus accruing the cost of providing incentives to communicate. On the other hand, it increases the information spread by influencers for fixed incentives.  When the first effect dominates, the seller should operate in a sparser network, somehow limiting the new demand attracted. Oppositely, if the latter demand effect prevails, maximizing the outreach should be the dominant concern of the seller.


The attractiveness of bonuses is enhanced by the presence of few very connected influencers (\textit{hubs}), who are cheaper to incentivize and guarantee a large information diffusion with little congestion. This is true for any given network density and drives the optimal choice of the seller. 
When operating in a denser network  gives access to super-hub consumers that are not present in a sparser one, a profit-maximizing company should always choose the latter. In contrast, if everyone was (homogeneously) more connected, then it might be optimal to sell the product in the sparser network where congestion is less severe.

However, any strategy that relies only on Public WOM will \emph{neglect} the fact that informed individuals can pass on product  information on multiple channels.  This raises scope for discriminating strategies between  publicly and privately generated product diffusion. In particular, considering also Private WOM induces the seller to offer a higher  public bonus  in order to move people on this  channel. This allows to optimally  spread information, not only relying on hubs but on a broader pool of informers. 



\bibliographystyle{chicago}
\bibliography{BibFile}

\newcommand{\SortNoop}[1]{}
\begin{thebibliography}{}

\bibitem[\protect\citeauthoryear{Acemoglu, Bimpikis, and Ozdaglar}{Acemoglu
  et~al.}{2014}]{AceBimp}
Acemoglu, D., K.~Bimpikis, and A.~Ozdaglar (2014).
\newblock {Dynamics of information exchange in endogenous social networks}.
\newblock {\em Theoretical Economics\/}~{\em 9\/}(1).

\bibitem[\protect\citeauthoryear{Akbarpour, Malladi, and Saberi}{Akbarpour
  et~al.}{2018}]{akbarpourjust}
Akbarpour, M., S.~Malladi, and A.~Saberi (2018).
\newblock Just a few seeds more: Value of network information for diffusion.

\bibitem[\protect\citeauthoryear{Albert, Jeong, and Barab{\'a}si}{Albert
  et~al.}{1999}]{albert1999internet}
Albert, R., H.~Jeong, and A.-L. Barab{\'a}si (1999).
\newblock Internet: Diameter of the world-wide web.
\newblock {\em Nature\/}~{\em 401\/}(6749), 130--131.

\bibitem[\protect\citeauthoryear{Arndt}{Arndt}{1967}]{arndt1967role}
Arndt, J. (1967).
\newblock Role of product-related conversations in the diffusion of a new
  product.
\newblock {\em Journal of Marketing Research\/}~{\em 4\/}(3), 291--295.

\bibitem[\protect\citeauthoryear{Ballester, Calv{\'o}-Armengol, and
  Zenou}{Ballester et~al.}{2006}]{ballester2006s}
Ballester, C., A.~Calv{\'o}-Armengol, and Y.~Zenou (2006).
\newblock Who's who in networks. wanted: The key player.
\newblock {\em Econometrica\/}~{\em 74\/}(5), 1403--1417.

\bibitem[\protect\citeauthoryear{Banerjee, Chandrasekhar, Duflo, and
  Jackson}{Banerjee et~al.}{2013}]{banerjee2013diffusion}
Banerjee, A., A.~G. Chandrasekhar, E.~Duflo, and M.~O. Jackson (2013).
\newblock The diffusion of microfinance.
\newblock {\em Science\/}~{\em 341\/}(6144), 1236498.

\bibitem[\protect\citeauthoryear{Banerji and Dutta}{Banerji and
  Dutta}{2009}]{dutta}
Banerji, A. and B.~Dutta (2009).
\newblock Local network externalities and market segmentation.
\newblock {\em International Journal of Industrial Organization\/}~{\em
  27\/}(5), 605--614.

\bibitem[\protect\citeauthoryear{Barab{\'a}si}{Barab{\'a}si}{2016}]{barabasi2016network}
Barab{\'a}si, A.-L. (2016).
\newblock {\em Network science}.
\newblock Cambridge university press.

\bibitem[\protect\citeauthoryear{Barab{\'a}si, Jeong, N{\'e}da, Ravasz,
  Schubert, and Vicsek}{Barab{\'a}si et~al.}{2002}]{barabasi2002evolution}
Barab{\'a}si, A.-L., H.~Jeong, Z.~N{\'e}da, E.~Ravasz, A.~Schubert, and
  T.~Vicsek (2002).
\newblock Evolution of the social network of scientific collaborations.
\newblock {\em Physica A: Statistical Mechanics and its Applications\/}~{\em
  311\/}(3), 590--614.

\bibitem[\protect\citeauthoryear{Berman}{Berman}{2016}]{berman2016referral}
Berman, B. (2016).
\newblock Referral marketing: Harnessing the power of your customers.
\newblock {\em Business Horizons\/}~{\em 59\/}(1), 19--28.

\bibitem[\protect\citeauthoryear{Bimpikis, Ozdaglar, and Yildiz}{Bimpikis
  et~al.}{2016}]{BimpOzd}
Bimpikis, K., A.~Ozdaglar, and E.~Yildiz (2016).
\newblock {Competitive Targeted Advertising Over Networks}.
\newblock {\em Operations Research\/}~{\em 64\/}(3), 705--720.

\bibitem[\protect\citeauthoryear{Biyalogorsky, Gerstner, and
  Libai}{Biyalogorsky et~al.}{2001}]{Biyalogorsky}
Biyalogorsky, E., E.~Gerstner, and B.~Libai (2001, August).
\newblock Customer referral management: Optimal reward programs.
\newblock {\em Marketing Science\/}~{\em 20\/}(1), 82--95.

\bibitem[\protect\citeauthoryear{Bloch}{Bloch}{2016}]{bloch2016handbook}
Bloch, F. (2016).
\newblock Targeting and pricing in social networks.
\newblock In A.~G. Yann~Bramoull\'e and B.~Rogers (Eds.), {\em The Oxford
  Handbook of the Economics of Networks}, pp.\  776--791. North-Holland:
  Amsterdam.

\bibitem[\protect\citeauthoryear{Bloch and Qu\'{e}rou}{Bloch and
  Qu\'{e}rou}{2013}]{bloch2011pricing}
Bloch, F. and N.~Qu\'{e}rou (2013).
\newblock Pricing in social networks.
\newblock {\em Games and Economic Behavior\/}~{\em 80\/}(C), 243--261.

\bibitem[\protect\citeauthoryear{Campbell}{Campbell}{2013}]{campbell2013word}
Campbell, A. (2013).
\newblock Word of mouth and percolation in social networks.
\newblock {\em American Economic Review\/}~{\em 103\/}(6), 2466--2498.

\bibitem[\protect\citeauthoryear{Campbell, Mayzlin, and Shin}{Campbell
  et~al.}{2017}]{campbell2017managing}
Campbell, A., D.~Mayzlin, and J.~Shin (2017).
\newblock Managing buzz.
\newblock {\em The RAND Journal of Economics\/}~{\em 48\/}(1), 203--229.

\bibitem[\protect\citeauthoryear{Candogan, Bimpikis, and Ozdaglar}{Candogan
  et~al.}{2012}]{candogan2012optimal}
Candogan, O., K.~Bimpikis, and A.~Ozdaglar (2012).
\newblock Optimal pricing in networks with externalities.
\newblock {\em Operations Research\/}~{\em 60\/}(4), 883--905.

\bibitem[\protect\citeauthoryear{Carroni, Pin, and Righi}{Carroni
  et~al.}{2017}]{carpinrighi2017}
Carroni, E., P.~Pin, and S.~Righi (2017).
\newblock Call your friend! {R}eal or {V}irtual?
\newblock \url{https://arxiv.org/abs/1710.08693}.

\bibitem[\protect\citeauthoryear{Chen, Zenou, Zhou, et~al.}{Chen
  et~al.}{2015}]{Zenou}
Chen, Y.-J., Y.~Zenou, J.~Zhou, et~al. (2015).
\newblock {\em Competitive pricing strategies in social networks}.
\newblock Centre for Economic Policy Research.

\bibitem[\protect\citeauthoryear{Cohen and Havlin}{Cohen and
  Havlin}{2003}]{cohen2003scale}
Cohen, R. and S.~Havlin (2003).
\newblock Scale-free networks are ultrasmall.
\newblock {\em Physical review letters\/}~{\em 90\/}(5), 058701.

\bibitem[\protect\citeauthoryear{Ebel, Mielsch, and Bornholdt}{Ebel
  et~al.}{2002}]{ebel2002scale}
Ebel, H., L.-I. Mielsch, and S.~Bornholdt (2002).
\newblock Scale-free topology of e-mail networks.
\newblock {\em Phys. Rev. E\/}~{\em 66}, 035103.

\bibitem[\protect\citeauthoryear{Erd{\H{o}}s and R{\'e}nyi}{Erd{\H{o}}s and
  R{\'e}nyi}{1959}]{erdHos1959random}
Erd{\H{o}}s, P. and A.~R{\'e}nyi (1959).
\newblock On random graphs.
\newblock {\em Publicationes Mathematicae Debrecen\/}~{\em 6}, 290--297.

\bibitem[\protect\citeauthoryear{Fainmesser and Galeotti}{Fainmesser and
  Galeotti}{2016}]{fainmesser2015pricing}
Fainmesser, I.~P. and A.~Galeotti (2016).
\newblock Pricing network effects.
\newblock {\em The Review of Economic Studies\/}~{\em 83\/}(1), 165--198.

\bibitem[\protect\citeauthoryear{Galeotti}{Galeotti}{2010}]{galeotti2010talking}
Galeotti, A. (2010).
\newblock Talking, searching, and pricing.
\newblock {\em International Economic Review\/}~{\em 51\/}(4), 1159--1174.

\bibitem[\protect\citeauthoryear{Galeotti and Goyal}{Galeotti and
  Goyal}{2009}]{galeotti2009influencing}
Galeotti, A. and S.~Goyal (2009).
\newblock Influencing the influencers: a theory of strategic diffusion.
\newblock {\em The RAND Journal of Economics\/}~{\em 40\/}(3), 509--532.

\bibitem[\protect\citeauthoryear{Golub and Sadler}{Golub and
  Sadler}{2016}]{learn2016handbook}
Golub, B. and E.~Sadler (2016).
\newblock Learning in social networks.
\newblock In A.~G. Yann~Bramoull\'e and B.~Rogers (Eds.), {\em The Oxford
  Handbook of the Economics of Networks}, pp.\  504--542. North-Holland:
  Amsterdam.

\bibitem[\protect\citeauthoryear{Goyal, Heidari, and Kearns}{Goyal
  et~al.}{2014}]{GOYAL2014}
Goyal, S., H.~Heidari, and M.~Kearns (2014).
\newblock Competitive contagion in networks.
\newblock {\em Games and Economic Behavior\/}.

\bibitem[\protect\citeauthoryear{Huston}{Huston}{2010}]{huston}
Huston, D. (2010).
\newblock
  \url{http://www.slideshare.net/gueste94e4c/dropbox-startup-lessons-learned-3836587}.

\bibitem[\protect\citeauthoryear{Iyengar, Van~den Bulte, and Valente}{Iyengar
  et~al.}{2011}]{iyengar2011opinion}
Iyengar, R., C.~Van~den Bulte, and T.~W. Valente (2011).
\newblock Opinion leadership and social contagion in new product diffusion.
\newblock {\em Marketing Science\/}~{\em 30\/}(2), 195--212.

\bibitem[\protect\citeauthoryear{Jackson, Rogers, and Zenou}{Jackson
  et~al.}{2017}]{jackson2016economic}
Jackson, M.~O., B.~W. Rogers, and Y.~Zenou (2017, March).
\newblock The economic consequences of social-network structure.

\bibitem[\protect\citeauthoryear{Kamada and {\"O}ry}{Kamada and
  {\"O}ry}{2017}]{kamada2017contracting}
Kamada, Y. and A.~{\"O}ry (2017).
\newblock Contracting with word-of-mouth management.

\bibitem[\protect\citeauthoryear{Katz and Lazarsfeld}{Katz and
  Lazarsfeld}{1966}]{lazarsfeld1955personal}
Katz, E. and P.~F. Lazarsfeld (1966).
\newblock {\em Personal Influence, The part played by people in the flow of
  mass communications}.
\newblock Transaction Publishers.

\bibitem[\protect\citeauthoryear{Katz and Shapiro}{Katz and
  Shapiro}{1985}]{katz1985network}
Katz, M.~L. and C.~Shapiro (1985).
\newblock Network externalities, competition, and compatibility.
\newblock {\em The American economic review\/}~{\em 75\/}(3), 424--440.

\bibitem[\protect\citeauthoryear{Lamberson}{Lamberson}{2016}]{diff2016handbook}
Lamberson, P. (2016).
\newblock Diffusion in networks.
\newblock In A.~G. Yann~Bramoull\'e and B.~Rogers (Eds.), {\em The Oxford
  Handbook of the Economics of Networks}, pp.\  479--503. North-Holland:
  Amsterdam.

\bibitem[\protect\citeauthoryear{Leduc, Jackson, and Johari}{Leduc
  et~al.}{2017}]{leduc2015pricing}
Leduc, M.~V., M.~O. Jackson, and R.~Johari (2017).
\newblock Pricing and referrals in diffusion on networks.
\newblock {\em Games and Economic Behavior\/}~{\em 104\/}(Supplement C), 568 --
  594.

\bibitem[\protect\citeauthoryear{Liljeros, Edling, Amaral, Stanley, and
  {\AA}berg}{Liljeros et~al.}{2001}]{liljeros2001web}
Liljeros, F., C.~R. Edling, L.~A.~N. Amaral, H.~E. Stanley, and Y.~{\AA}berg
  (2001).
\newblock The web of human sexual contacts.
\newblock {\em Nature\/}~{\em 411\/}(6840), 907--908.

\bibitem[\protect\citeauthoryear{Lobel, Sadler, and Varshney}{Lobel
  et~al.}{2016}]{lobel2016customer}
Lobel, I., E.~Sadler, and L.~R. Varshney (2016).
\newblock Customer referral incentives and social media.
\newblock {\em Management Science\/}~{\em 63\/}(10), 3514--3529.

\bibitem[\protect\citeauthoryear{L{\'o}pez-Pintado}{L{\'o}pez-Pintado}{2008}]{lopez2008diffusion}
L{\'o}pez-Pintado, D. (2008).
\newblock Diffusion in complex social networks.
\newblock {\em Games and Economic Behavior\/}~{\em 62\/}(2), 573--590.

\bibitem[\protect\citeauthoryear{L{\'o}pez-Pintado}{L{\'o}pez-Pintado}{2012}]{lopez2012influence}
L{\'o}pez-Pintado, D. (2012).
\newblock Influence networks.
\newblock {\em Games and Economic Behavior\/}~{\em 75\/}(2), 776--787.

\bibitem[\protect\citeauthoryear{Pin and Rogers}{Pin and
  Rogers}{2016}]{random2016handbook}
Pin, P. and B.~Rogers (2016).
\newblock Stochastic network formation and homophily.
\newblock In A.~G. Yann~Bramoull\'e and B.~Rogers (Eds.), {\em The Oxford
  Handbook of the Economics of Networks}, pp.\  138--166. North-Holland:
  Amsterdam.

\bibitem[\protect\citeauthoryear{Shi}{Shi}{2003}]{shi2003social}
Shi, M. (2003).
\newblock Social network-based discriminatory pricing strategy.
\newblock {\em Marketing Letters\/}~{\em 14\/}(4), 239--256.

\bibitem[\protect\citeauthoryear{Sundararajan}{Sundararajan}{2007}]{Sundararajan}
Sundararajan, A. (2007).
\newblock Local network effects and complex network structure.
\newblock {\em The BE Journal of Theoretical Economics\/}~{\em 7\/}(1),
  1935--1704.

\bibitem[\protect\citeauthoryear{Ugander, Karrer, Backstrom, and
  Marlow}{Ugander et~al.}{2011}]{ugander2011anatomy}
Ugander, J., B.~Karrer, L.~Backstrom, and C.~Marlow (2011).
\newblock The anatomy of the facebook social graph.

\bibitem[\protect\citeauthoryear{Van~den Bulte and Joshi}{Van~den Bulte and
  Joshi}{2007}]{van2007new}
Van~den Bulte, C. and Y.~V. Joshi (2007).
\newblock New product diffusion with influentials and imitators.
\newblock {\em Marketing Science\/}~{\em 26\/}(3), 400--421.

\bibitem[\protect\citeauthoryear{Watts and Strogatz}{Watts and
  Strogatz}{1998}]{watts1998collective}
Watts, D.~J. and S.~H. Strogatz (1998).
\newblock Collective dynamics of 'small-world' networks.
\newblock {\em nature\/}~{\em 393\/}(6684), 440.

\bibitem[\protect\citeauthoryear{Yu and Van~de Sompel}{Yu and Van~de
  Sompel}{1965}]{yu1965networks}
Yu, P. and H.~Van~de Sompel (1965).
\newblock Networks of scientific papers.
\newblock {\em Science\/}~{\em 169}, 510--515.

\end{thebibliography}

\setcounter{section}{1}

\appendix

\global\long\def\thesection{Appendix \Alph{section}}
 \global\long\def\thesubsection{\Alph{section}.\arabic{subsection}}
 \setcounter{proposition}{0} \global\long\def\theproposition{\Alph{proposition}}
 \setcounter{definition}{0} \global\long\def\thedefinition{\Alph{definition}}

\section{}
\label{app:proofs}

\subsection*{Proof of Proposition \ref{prop:all1/2}, page \pageref{prop:all1/2}}

The profit is concave in $p_{2}$.
So, the optimal $p_2^*$ solves the first-order condition:
\begin{eqnarray}
1-2p_2 & = & \frac{d}{d p_2} \left(  b (1-p_2) \right) \nonumber \\
& = &
-b + (1-p_2) \frac{\partial b}{\partial \phi} \frac{d \phi}{d p_2} \nonumber \\
& = & 
-b + \underbrace{(1-p_2) \left(  - \frac{c}{\phi^2 \underline{k}} \right)
\left(  - \sum_{k=1}^{\infty} g(k)  \left( \frac{1-(1-L)^k}{k L} \right) \right) }_{= \frac{b}{\phi} \cdot \phi} \ \ , \nonumber
\end{eqnarray}
but RHS is just $0$, so that $p_2^* = \frac{1}{2}$, independently on any other variable.
Now, instead of problem \eqref{eq:monopolist_problem} we can address problem \eqref{eq:monopolist_problem3}, and look for optimal $\underline{k}$ instead of optimal $b$.
 Plugging the optimal price into the profit function, the seller  has to solve: 
\begin{eqnarray}
\max\limits_{\underline{k}}\pi(1/2,\underline{k})  &=& 
\max\limits_{\underline{k}} \left[ (1-\beta)  + \beta  \left(\frac{1}{2}-\frac{c}{\underline{k} \phi (L(\underline{k}),1/2)}\right) \frac{ \Gamma(L(\underline{k}))}{2} \right] \nonumber    \ \ ,\end{eqnarray}
which has the same maximum as
\begin{eqnarray*}
\max\limits_{\underline{k}}\left[\frac{\Gamma(L(\underline{k}))}{4}-\frac{c\Gamma(\underline{k},1)}{2\underline{k} \phi ( L(\underline{k}),1/2)}\right]  \ \ .\end{eqnarray*}
The objective function above is lower semicontinuous and has a limited number of jumps. Therefore, it always exists some $\underline{k}^{*}$ that maximizes it.  \qed

\subsection*{Proof of Proposition \ref{prop:cost_per_contact}, page \pageref{prop:cost_per_contact} }

Similarly to the case of lump-sum cost,  independently of the choice of $b$, charging a  price  $p_{1}=1$  to informed consumers and a price $p_{2}=1/2$ are the optimal strategy. 
Plugging the optimal prices into the profit function, the seller sets $L^{*}$  to solve: 
\begin{eqnarray*}
\max\limits_{L}\pi(L)  &=&
\max\limits_{L}\left[(1-\beta) + \frac{\beta}{2}  \left(\frac{1}{2}-\frac{ c_{pr}}{\phi(L,1,1/2)}\right) \Gamma(1/2,L)\right]\\
&=&\max\limits_{L }  \left(\frac{1}{2}-\frac{ c_{pr}}{\phi(L,1,1/2)}\right) \Gamma(1/2,L)
\end{eqnarray*}

The first order condition for this problem is:

$$\frac{\partial \pi}{\partial L}=\frac{\partial \Gamma}{\partial L}\left(\frac{1}{2}-\frac{c_{pr}}{\phi}\right)+\frac{c_{pr} \Gamma}{\phi^{2}}\frac{\partial \phi}{\partial L}=0, $$

Notice that, provided that $1/2>c_{pr}/\phi$,\footnote{This is the condition needed for the referral program to be profitable. If this condition is not met, the bonus exceeds the price. } profits are concave in $L$. Indeed: 
$$\frac{\partial^{2} \pi}{\partial^{2} L }=\underbrace{\frac{\partial^{2} \Gamma}{\partial^{2} L }\frac{1}{2}-\frac{\partial^{2} \Gamma}{\partial^{2} L }\frac{c_{pr}}{\phi} }_{<0\text{ when }1/2>c_{pr}/\phi}+\underbrace{c_{pr}\left[\frac{\frac{\partial\Gamma}{\partial L }\frac{\partial^{2} \phi}{\partial^{2}  L }}{\phi^{2}}+\frac{\left(\frac{\partial\Gamma}{\partial L}\frac{\partial\phi}{\partial L}-\Gamma\frac{\partial^{2} \phi}{\partial^{2}  L }\right)-2\phi\Gamma\left(\frac{\partial\phi}{\partial L}\right)^{2}}{\phi^{3}}\right]}_{<0\text{ by concavity of $\Gamma$ and convexity of $\phi$}}<0,$$

Now, if $ L  \rightarrow 0$, $\frac{\partial \Gamma}{\partial L}=\sum\limits_{k=0}^{\infty}g(k)\frac{k }{ 2}$, $\frac{\partial \phi}{\partial L}=\sum\limits_{k=0}^{\infty}g(k)\frac{1-k}{8}$, $\phi=1/2$ and $\Gamma=0$. Therefore:

$$\frac{\partial \pi}{\partial L}|_{ L =0}=\left(\frac{1}{2}-2c_{pr}\right)\sum\limits_{k=0}^{\infty}g(k)\frac{k }{ 2}>0\text{ when }c_{pr}<1/4.$$

Moreover, the fact that   $\frac{\partial \Gamma}{\partial L}|_{L=1}=0$ together with  $\frac{\partial \phi}{\partial L}<0$ imply that  the profit is decreasing in $L$ when $L=1$. As a consequence $L^{*}$ is always interior, it is unique and it solves: 

$$\frac{\partial \Gamma}{\partial L}\left(\frac{1}{2}-\frac{c_{pr}}{\phi}\right)+\frac{c_{pr} \Gamma}{\phi^{2}}\frac{\partial \phi}{\partial L}=0.$$
\qed

\subsection*{Proof of Corollary \ref{penepene}, page \pageref{penepene}}

Under public communication, the profit of the seller is \begin{eqnarray*}
(1-\beta)  + \beta  \left(\frac{1}{2}-\frac{c}{\underline{k} \phi (L(\underline{k}),1/2)}\right) \frac{ \Gamma(L(\underline{k})) }{2}
\end{eqnarray*}
When $\underline{k}$ is set equal to $k_{f\min}$, $L=1$, so that $\Gamma=1$ is clearly maximal, as every consumer willing to pay the price of the product will buy the product. In this case 
\[
\phi(L,1/2) = \frac{1}{2} \sum_{k=1}^{\infty} g(k)  \left( \frac{1-(1-L)^k}{k L} \right) = \frac{1}{2}\sum_{k=1}^{\infty} g(k) \left( \frac{1-0}{k} \right) =   E_k [ 1/k ]/2 \ \ .
\]
Pugging $\phi(1)=E_k [ 1/k ]/2$ into the profit, it is easy to verify that it is fulfilled only  if $E_k [ 1/k ] >  \frac{4 c}{k_{f \min}(4-3\beta)}$. 
  Similarly, $\phi(1,1/2)=E_k [ 1/k ]$ also under Private WOM, where the profit is: 
\begin{eqnarray*}
(1-\beta)  + \beta  \left(\frac{1}{2}-\frac{c_{pr}}{ \phi (L,1/2)}\right) \frac{ \Gamma(L) }{2}=(1-\beta)  + \beta  \left(\frac{1}{2}-\frac{c_{pr}}{ E_k [ 1/k ]/2}\right) \frac{1}{2}
\end{eqnarray*}
which is positive if  $E_k [ 1/k ] >   \frac{4 \beta }{4-3\beta}c_{pr}  $.

\qed

\subsection*{Proof of Corollary \ref{prop:misalignment}, page \pageref{prop:misalignment}}

Since  the choice of   $\underline{k}$ is concerned in optimizing \eqref{eq:monopolist_problem3},  we  refer only to  the profit the  seller makes on second-period consumers, expressing $b$ in terms of the other variables:  
\begin{equation}
\label{profit_piece}
\pi_2(1/2, \underline{k} )  = \frac{\beta}{2} \left( \frac{1}{2}-\frac{c}{\phi \underline{k}} \right)\Gamma(L)\ \ .
\end{equation}
The seller would maximize profits and diffusion simultaneously  if $\underline{k}=k_{f \min}$ and then $L=1$. 
In this case $\phi(L) = E_k [ 1/k ]$. 
However, if $\phi(L) <  \frac{2c}{k_{f \min}}$, then $\frac{1}{2}-\frac{c}{\phi k_{f \min}} <0 $, and so the part of profits from (\ref{profit_piece}) is negative. 
If instead the  seller chooses $\underline{k} > k_{f \max}$, then $L=0$, and the part of profits from (\ref{profit_piece}) is null.
So,  $\underline{k}=k_{f \min}$ cannot be an optimal choice for the seller, because it is dominated by $\underline{k}=k_{f \max}$. \qed

\subsection*{Proof of Lemma \ref{FOSDWelfare}, page \pageref{FOSDWelfare}}

Since the threshold $\underline{k}$ remains fixed, a  FOSD shift from  $f(k)$ to $f'(k)$   makes 
\[
L = \sum_{k=\underline{k}}^{\infty} f(k) \leq \sum_{k=\underline{k}}^{\infty} f'(k) = L'
\ \ .
\]
Therefore, each $1-(1-L)^k$ increases. 
In the expression of $\Gamma(L')$, as given by equation \eqref{eq:Gamma},
$1-(1-L)^k$ is also an increasing function in $k$.
So, given that also $g'(k)$ FOS-dominates $g(k)$, we can use Definition \ref{def_fosd} and obtain
\[
\Gamma(L) = \sum_{k=1}^{\infty} g(k) \left(  1-(1-L)^k \right) \leq \sum_{k=1}^{\infty} g(k) \left(  1-(1-L')^k \right) \leq \sum_{k=1}^{\infty} g'(k) \left(  1-(1-L')^k \right) = \Gamma (L')
.\]
So 
$ \Gamma(L) \leq \Gamma (L') $. \qed

\subsection*{Proof of Lemma \ref{prop:FOSD_homogeneous}, page \pageref{prop:FOSD_homogeneous}}\label{FOSD_homogeneous}

Let us consider a network where all consumers have  in-degree equal to $k_g$ and out-degree equal to $k_f$.  From the consistency condition in \eqref{eq:consistency} it must hold that $(1-\beta)k_f=\beta k_g$.  Plugging into   \eqref{eq:Gamma} and \eqref{eq:phi}, we get:

\begin{equation}
\label{system_hom_degree}
\begin{array}{c}
\Gamma(L)=1-(1-L)^{k_g},\\
\phi(L, k_g)=\frac{1}{L k_g}\left(1-(1-L)^{k_g}\right)=\frac{1}{L k_g}\Gamma(L).
\end{array}
\end{equation}
Using consistency condition \eqref{eq:consistency}, the profit in equation \eqref{profit_piece} becomes either null (if the seller does not use word-of-mouth)\footnote{ This will be the case whenever $\frac{c}{\phi(L)k_{f}}<1/2$ for all $L\in(0,1)$, so that profits made on uninformed consumers would  be negative.  } or: 
\begin{equation}
 \frac{\beta}{2} \left( \frac{1}{2}-  \frac{1-\beta}{\beta} \frac{L c }{\Gamma(L)} \right)\Gamma(L)
= \frac{\beta}{4} \Gamma(L) -  L (1-\beta) c   \ \ . 
\label{simple_objective}
\end{equation}
Notice that equation \eqref{simple_objective} has a concave increasing part and a linearly decreasing one.
A FOSD shift would affect the former but not the latter. On the one hand,  in a denser network the function $\Gamma'(L)$ maps $L$ into higher values than the original $\Gamma(L)$. On the other hand, the FOSD shift does not entail any effect on the linear part, which depends only on the slope $(1-\beta) c$ that is not affected by network density. Therefore, for any $L$ the after-FOSD profit will always be above than the pre-FOSD one. 
Both information diffusion   and seller's profit increase.
\eproof

\subsection*{Proof of Lemma \ref{prop:INHOMOWELF}, page \pageref{prop:INHOMOWELF}}\label{INHOMOWELF}

If the seller chooses some $L^{*}$, information diffusion  only  depends  on 
\[
\Gamma(k_g) = 1-(1-L^{\ast} )^{k_g} \ \ .
\]
When $k_g$ increases to some $k'_g$, $L^*$ may move to some $L'=L^*-\epsilon$ with $\epsilon>0$. Plugging into $\Gamma^{\prime}$, we get that the denser network   enhances market diffusion if:

\begin{eqnarray}
\Gamma'>\Gamma
& \Leftrightarrow &
(1-(L^{\ast}-\epsilon) )^{k_g^{\prime}}<(1-L^{\ast} )^{k_g}\nonumber  \\
& \Leftrightarrow & k'_g >\frac{\log[1-L^{\ast} ]}{\log[1-L^{\ast}+\epsilon]}k_g  \nonumber  \\
& \Leftrightarrow & 
\frac{k'_g}{k_g}>\frac{\log[1-L^{\ast} ]}{\log[1-L^{\ast}+\epsilon]} \ \ .
\label{SuffCondFOSDwelfareenh}
\end{eqnarray}

Condition in \eqref{SuffCondFOSDwelfareenh} is sufficient for the FOSD shift to improve the proportion of people receiving the information. Clearly, if $L^{*}=1$, $\Gamma^{\prime}\leq\Gamma$. When $L^{\ast}=1-\delta$ with $\delta>0$ but very small, the condition becomes $\frac{k'_g}{k_g}>\frac{\log[\delta ]}{\log[\delta+\epsilon]}$.

\eproof

\subsection*{Proof of Proposition \ref{prop:HOMOIN}, page \pageref{prop:HOMOIN} }\label{HOMOIN}

Let us consider a network where all consumers have  in-degree equal to $k_g$.  
Let us first consider what would be the profit of the seller choosing a certain $L$.
Plugging  into \eqref{eq:Gamma} and \eqref{eq:phi}, we get again the system from 
\eqref{system_hom_degree}.
Now, however, $\underline{k}$ will vary as a certain  $L$ is targeted, and we have that the profit in equation \eqref{profit_piece} becomes either null (if the  seller does not use word of mouth) or:
\begin{equation}
 \frac{\beta}{2} \left( \frac{1}{2}-  \frac{ L k_g}{\underline{k}} \frac{c}{\Gamma(L)} \right)\Gamma(L)
= \frac{\beta}{4} \Gamma(L) -  c \frac{L k_g}{\underline{k}}  \ \ . \nonumber
\end{equation}

Now suppose to fix both $\underline{k}$ and the fraction $q$ of consumers with degree $\underline{k}$ that pass the information in equilibrium.
We have that:
\[
L = q f(\underline{k}) \cdot \underline{k} + \sum_{k=\underline{k}+1}^{k_{f \max} } f(k) \cdot k \ \ .
\]
Any couple $(\underline{k},q)$ will be feasible also going to a denser network, and fixing them we will have $L'>L$.\\
For any couple $(\underline{k},q)$, the profit in equation \eqref{profit_piece} becomes: 
\begin{eqnarray}
\pi (\underline{k},q) = \frac{\beta}{4} \Gamma(L)-    c \frac{k_g L}{\underline{k}} \ \ .
\label{profit_piece_lattice}
\end{eqnarray}

Now consider the case in which $\underline{k}=k_{f \min}$ (minimal cost for the  seller to provide incentives to communication) and $\Gamma\rightarrow1$ (maximal information  spread). A sufficient condition for each couple $(\underline{k},q)$ to provide negative profits is that \begin{revs} we have negative profits even in the best scenario.   This corresponds formally to:\end{revs}\footnote{Notice that, in this simple case, we would have $\rho=f(k_{f \max})$, i.e. only more connected influencers pass the information and maximal information spread is guaranteed.}  
 
\[
c > \frac{\beta}{4} \frac{k_{f \max}}{k_g f(k_{f \max})} \ \ ,
\]
that using condition \eqref{eq:consistency}, becomes

\[
c > \frac{\beta^{2}}{4} \frac{k_{f \max}}{(1-\beta)E_f (k) f(k_{f \max})} \ \ ,
\]

Going to a denser network we can get $\frac{k'_{f \max}}{ E_f (k') }$ as close to $1$ as we want, so that $f(k_{f \max})$ as well goes to $1$. Therefore,  
$c > \frac{\beta^2}{4(1-\beta) }$ becomes a sufficient condition for the existence of a denser network that is not profitable for the seller.
Notice that if $k_{f \max}$ was high enough in the original network, then it could have been possible to find a couple $(\underline{k},q)$  such that \eqref{profit_piece_lattice} was positive in the original sparser network, but negative in the denser network.
\eproof

\subsection*{Proof of Proposition \ref{prop:general} page \pageref{prop:general}}\label{proof:general}

Suppose to create \emph{super hubs} in the $f$ distribution, and to increase uniformly the $g$ distribution.
Then, to respect the network constraint, we give degree $\bar{k}>k_{f \max}$ to the hubs, and add $m$ links to the uninformed consumers, such that
\begin{eqnarray}
(1-f(\bar{k})) \left( \sum f(k) k \right) + f(\bar{k} ) \bar k & = & \frac{\beta}{1-\beta} \left( \sum g(k) (k+m) \right)  \nonumber \\
& = & \left( \sum f(k) k \right) + m \frac{\beta}{1-\beta} 
\ \ , \label{consistencyFOSD}
\end{eqnarray}
which implies
\[
f(\bar{k} ) \bar{ k} =  m \frac{\beta}{1-\beta} +  f(\bar{k} ) \left( \sum f(k) k \right)  \ \ .
\]
It is important to consider that all these expressions for $m$ make sense even if  the latter is a real non-integer number, and we assume to give the integer part $\lfloor m \rfloor$ to each uninformed consumer and the remaining real part $m-\lfloor m \rfloor$ is a probability for each of them to receive the $\lfloor m \rfloor+1^{th}$ link. That is because equation \eqref{consistencyFOSD} is linear in $m$.

\bigskip

\noindent {\bf Step 1: for any $\epsilon>0$, we can get $\Gamma - \epsilon < \Gamma' < \Gamma$.} \\
The old $\Gamma$ was
\[
\Gamma= 1- \left( \sum g(k) \left(  1-L^{*}\right)^k \right)  \ \ .
\]
The new $\Gamma'$ will become
\[
\Gamma'= 1- \left( \sum g(k) \left(  1-L' \right)^k \right) \left(  1-L' \right)^m \ \ .
\]

Since $\lim_{m\rightarrow 0} \Gamma'=0$ and $\lim_{m\rightarrow \infty} \Gamma'=1$, we can pick arbitrarily any value for $\Gamma'$, keeping also $\bar{k}$ free to move as high as wanted (since we have still freedom on $f(\bar{k})$).
So, for each $\epsilon>0$, we can take 
\[
\Gamma-\epsilon < \Gamma' < \Gamma \ \  ,
\]
and are still free to pick any $\bar{k}$.

\bigskip

\noindent {\bf Step 2: we can improve on the payoff from the old network.} \\
Now, the seller's objective function can be reduced to
\begin{equation}\label{eq:monopolist_problem4}
\left( \frac{1}{2} -   \frac{c}{\underline{k} \phi \left( \underline{k}, \frac{1}{2} \right)  } \right)  \Gamma(\underline{k} ) \ \ , \end{equation}
and in this case we compare
\[
\left( \frac{1}{2} -   \frac{c}{\underline{k}^{*} \phi \left( \underline{k}^{*}, \frac{1}{2} \right)  } \right)  \Gamma
\mbox{ \ \ with \ \ } 
\left( \frac{1}{2} -   \frac{c}{ \bar{k} \phi' \left( \frac{\rho'}{2} , \frac{1}{2} \right) } \right)  (\Gamma  - \epsilon ) \ \  .
\]
As we can take $f(\bar{k} ) \rightarrow 0 $ and $ \bar{k} \rightarrow \infty $, let us show that we can achieve 
\[
\lim_{\bar{k} \rightarrow \infty} \bar{k} \phi' \left(\bar{k}, \frac{1}{2} \right) = + \infty \ \ .
\]
This is always possible if we consider that $\phi' \left( L^{\prime} , \frac{1}{2} \right)$ is bounded below by
\[
E' (1/k) = \sum g'(k)/k =  (1-f(\bar{k}) ) E (1/k)  + \left(   m \frac{\beta}{1-\beta} +  f(\bar{k} ) \left( \sum f(k) k \right) \right)  \frac{1}{\bar{k}^2} \ \ .
\]
Since $\Gamma<1$, we can set $\epsilon <  \frac{c}{ k_{f \min} \phi \left( \frac{1}{2}, \frac{1}{2} \right)}$, and we can set a  $ \bar{k}$ high enough so that 
\[
\left( \frac{1}{2} -   \frac{c}{\underline{k}^{*}\phi \left( \underline{k}^{*}, \frac{1}{2} \right)  } \right)  \Gamma
<
\left( \frac{1}{2} -   \frac{c}{ \bar{k} \phi' \left(\bar{k} , \frac{1}{2} \right) } \right)  ( \Gamma  - \epsilon ) \ \  .
\]
\bigskip

\noindent {\bf Step 3: in the new network, the  seller will not choose a $\underline{k}<\bar{k}$ .} \\
Consider again the  seller problem in equation \eqref{eq:monopolist_problem4}.
We have seen that we can set 
\[
\lim_{\bar{k} \rightarrow \infty} \bar{k} \phi' \left(\bar{k} , \frac{1}{2} \right) = + \infty \ \ .
\]
It is not difficult to see that all the levels of $\underline{k}$ that were feasible in the old network will be feasible also in the new network, but at a higher cost.
That is because $\underline{k}$ will be the same but expected gains from $\phi$ will  be decreased by the congestion created by the new super hubs.
So, suppose that the  seller chooses some $L''> L'$, then we can still maintain the following inequalities:
\[
\left( \frac{1}{2} -   \frac{c}{ \Lambda(L^{\prime\prime}) \phi \left(L^{\prime\prime}, \frac{1}{2} \right)  } \right)  
<
\left( \frac{1}{2} -   \frac{c}{ \Lambda(\rho^*) \phi \left( \frac{\rho^*}{2}, \frac{1}{2} \right)  } \right) 
< 
\left( \frac{1}{2} -   \frac{c}{ \bar{k} \phi' \left(\bar{k} , \frac{1}{2} \right) } \right)  ( 1 - \delta - \epsilon ) \ \  ,
\]
where $\Lambda(L^{\prime\prime})=\left\{\underline{k}:\sum\limits_{k=\underline{k}}^{\infty}f(k)=L^{\prime\prime}\right\}$.
The last step is guaranteed by the assumption that $\delta  <  \frac{2 c}{ \underline{k}^* \phi \left(\underline{k}, \frac{1}{2} \right)}$, and by the fact that we can achieve 
$\epsilon \rightarrow 0$ and 
\[
\lim_{\bar{k} \rightarrow \infty} \bar{k} \phi' \left( \bar{k} , \frac{1}{2} \right) = + \infty \ \ .
\]
So for any of the feasible $\underline{k}$ in the old network, we can choose $\epsilon$, $\bar{k}$ and $m$, so that the  seller can do better in the new network choosing to do word-of-mouth only with  the superhubs.
Since, the set of all feasible $\underline{k}$'s in the old network is finite, we can satisfy the conditions of this last step of the proof.
\eproof

\subsection*{Proof of Proposition \ref{prop:allmix}, page \pageref{prop:allmix}}

The profit from \eqref{eq:monopolist_problem_mix}  is concave in $p_{2}$.
So, the optimal $p_2^*$ solves the first-order condition:
\begin{eqnarray}
1-2p_2 & = & \frac{d}{d p_2} \left(  \mathbb{E}[B] (1-p_2) \right) \nonumber \\
& = &
-\mathbb{E}[B] + (1-p_2) \frac{\partial \mathbb{E}[B]}{\partial \phi} \frac{d \phi}{d p_2} \nonumber \\
& = & 
-\mathbb{E}[B] + \underbrace{(1-p_2) \left(  - \frac{c}{\phi^2 } \right)
\left(  - \sum_{k=\underline{k}}^{\infty}f(k)\left[\frac{1+\underline{k}}{2\underline{k}}+\sum\limits_{k=k\min}^{\underline{k}}f(k)\right]\right)}_{\frac{\mathbb{E}[B]}{\phi}\phi } \ \ , \nonumber
\end{eqnarray}
but RHS is just $0$, so that $p_2^* = \frac{1}{2}$, independently on any other variable.
 Plugging the optimal price into the profit function, the seller  has to solve: 
\begin{equation*}
\max\limits_{\underline{k}}\pi(1/2,\underline{k})  = 
\max\limits_{\underline{k}}\left[(1-\beta)  + \frac{\beta}{2}
\left(\frac{1}{2}-\mathbb{E}[B|\underline{k}]\right)\Gamma(L(\underline{k}))\right]
\end{equation*}
which has the same maximum as
\begin{eqnarray*}
\max\limits_{\underline{k}}\pi(1/2,\underline{k})  = 
\max\limits_{\underline{k}}\left[\left(\frac{1}{2}-\mathbb{E}[B|\underline{k}]\right) \frac{\Gamma(L(\underline{k})}{2}\right]
.\end{eqnarray*}
The objective function above is lower semicontinuous and has a limited number of jumps. Therefore, it always exists some $\underline{k}^{*}$ that maximizes it.  \qed

\section{Construction of the network}
\label{Construction}

We provide here a brief mathematical description of our approach in defining the network.
We adopt two  continua of consumers, the \emph{informed} ones and the \emph{uninformed} ones.
They can be thought as two segments of the real line: $[0,1-\beta)$ and $[0,\beta]$.
We consider a (possibly infinite) discrete series $f(k)$ such that $\sum f(k) = 1$ and we assume that a fraction $f(k) (1-\beta)$ of informed consumers are connected to $k$ point in the segment of uninformed consumers.
Formally, we have a function from a segment of size $f(k) (1-\beta)$ to $[0,\beta]^k$.
On top of that we impose two constraints:
First, we assume also that the  in-degree distribution is ruled by a  (possibly infinite) discrete series $g(k)$, such that $\sum g(k) = 1$.  So, a mass $g(k) \beta$ of them have in-degree $k$.
Second, we impose the balance condition from equation \eqref{eq:consistency}.
In this way, we restrict on the set of all possible countable functions from the segments of size $f(k) (1-\beta)$ to $[0,\beta]^k$ that satisfy the addional constraints imposed by $g(k) $ and by the balance condition.

\subsection{The networks in the numerical studies  in Sections \ref{profitmax} and \ref{mix}.}

For our  analysis, we   use two classes of networks,  which represent stylized versions of empirically observed social networks (see \citealt{random2016handbook} for a survey related to the economic literature).
The first type is the \cite{erdHos1959random} random network also known as  Bernoulli random network. These graphs are characterized by a given number of nodes $n$ and a given probability $0\leq \lambda \leq 1$, which describes the chance of each link between pairs of nodes to exist. When  $\lambda$ is assumed to be equal for each pair of nodes, these networks are characterized by a binomial distribution of degrees, i.e.,  $\forall k\in[1, n-1]$:
\begin{equation*}
h(k)={n - 1 \choose k}\lambda^k (1-\lambda)^{n-1-k},
 \label{binomial}
\end{equation*}
where $\lambda n$ approximates the characteristic degree of nodes in the networks, and most of the nodes have a degree close to it. In other terms, $\lambda$ can be considered as a measure of network density. 

The second type of degree distribution characterizes networks defined as scale-free due to the tendency of second and higher order moments to diverge. This type of construction does fit many of the characteristics of empirical social networks, in particular the observation that a lot of them approximately follow a power-law degree distribution (for specific examples see \citealt{ugander2011anatomy,ebel2002scale,liljeros2001web,barabasi2002evolution,yu1965networks,albert1999internet}).
  Formally, we study networks with degrees up to $N$ and a degree distribution given by:
\begin{equation*}
h(k)=\frac{1/k^{\gamma}}{\sum\limits_{n\in N}(1/n^{\gamma})},
\end{equation*}
where $2< \gamma \leq 3$ represents the slope of the power law.\footnote{In between these two values the first moment of the degree distribution is finite, but the second and higher moments diverge as the network size becomes infinite. The boundaries are justified by the fact that most empirically observed social networks exhibit a slope between these two values, which implies a ultra small-world network (\citealt{cohen2003scale}).  Notice that increasing the parameter $\gamma$ implies lowering the probabilities to observe highly-connected individuals, thus leading to sparser networks, and to smaller hubs.}

Scale-free networks entail the presence of hubs, i.e., nodes with very high degree with respect to the network' average, while in random networks these disproportionally-connected  nodes  are extremely rare. For this reason, in what follows we refer to the first type of networks as {\it networks with hubs} and to the latter as {\it networks without hubs}.  
Given that the only topological characteristic of the networks which is relevant for the definition of participants decisions in our model is the distribution of degrees, our numerical analysis of random and scale-free networks completes the study of the effects of in- and out- degree distributions on profits and information diffusion for what concerns empirically-observed network structures.

Considering these network structures, FOSD shifts in Section \ref{profitmax} result from increasing $\lambda_{g}$ (for the random network) and decreasing $\gamma_{g}$ (for the scale-free networks) by $2.5\cdot 10^{-2}$, without changing $k_{f \max}$ and $k_{g \max}$. In both cases we are thus making each possible link between arbitrary agents $i$ and $j$ more likely to exist.

\end{document}